\documentclass{kapproc} 

\usepackage{epsfig}
\let\footnote\savefootnote

\let\footnoterule\savefootnoterule

\normallatexbib
\begin{document}


\articletitle[]{THE STANDARD MODEL\\AND THE TOP QUARK}
\author{Scott Willenbrock}
\affil{Physics Department\\University of Illinois at Urbana-Champaign\\1110
W.~Green St., Urbana, IL   61801} \email{willen@uiuc.edu}

\begin{abstract}

The top quark is one of the least well-studied components of the standard
model.  In these lectures I discuss the expected properties of the top quark,
which will be tested at the Fermilab Tevatron and the CERN Large Hadron
Collider.  I begin with a modern review of the standard model, emphasizing the
underlying concepts. I then discuss the role the top quark plays in precision
electroweak analyses.  The last two lectures address the strong and weak
interactions of the top quark, with an emphasis on the top-quark spin.

\end{abstract}

The top quark is the least well-studied of the quarks.  Why is the top quark an
interesting and worthwhile object to study? Here are four of the most
compelling reasons:
\begin{itemize}
\item A more accurate measurement of the top-quark mass is valuable as an input
to precision electroweak analyses.
\item We would like to know if the top quark is just an ordinary quark, or if
it is exotic in some way.
\item The top quark may be useful to discover new particles.  For example, of
all the fermions, the Higgs boson couples most strongly to the top quark.  It
might be possible to observe the Higgs boson produced in association with a
$t\bar t$ pair.
\item Events containing top quarks are backgrounds to new physics that we hope
to discover.  This may sound mundane, but it is extremely important.  For
example, the discovery of the top quark itself was only possible once we
understood the background from $W+$jets.
\end{itemize}

Although these lectures are principally about the top quark, I have chosen to
broaden them to include a review of the standard model.  The top quark is very
much a part of the standard model, and it is useful to discuss the physics of
the top quark from that perspective.  The physics of the top quark is a vast
subject, and cannot be covered in a few lectures.  Instead, I have chosen
several subjects of broad interest related to the top quark, and discuss them
in some depth, with an emphasis on the underlying concepts.  I have also
included several exercises with each lecture, which I strongly urge you to
perform.  They will engage you with the material in a way that will help
solidify your understanding.  The exercises can be performed using only
material contained in these lectures.  The exercises are of various levels of
difficulty, indicated by $\ast$ (easy), $\ast\ast$ (moderate), and
$\ast\ast\ast$ (hard). Solutions are provided in an appendix.

The first lecture is a review of the standard model from a modern point of
view.  It assumes the reader already has some familiarity with the standard
model, and concentrates on the concepts that underlie the theory.  The second
lecture discusses the role the top quark plays in precision electroweak
analyses via one-loop processes. The third and fourth lectures discuss the
strong and weak interactions of the top quark, respectively, with an emphasis
on the top-quark spin.

\section{The Standard Model}\label{sec:sm}

In Table~\ref{tab:sm} I list the fermion fields that make up the standard
model, along with their $SU(3)\times SU(2)\times U(1)_Y$ quantum numbers.  The
index $i=1,2,3$ on each field refers to the generation, and the subscript $L,R$
refers to the chirality of the field ($\psi_{L,R}\equiv {1\over 2}
(1\mp\gamma_5)\psi$). The left-chiral and right-chiral fields corresponding to
a given particle have different $SU(2)\times U(1)$ quantum numbers, which
leads to parity violation in the weak interaction.

\begin{table}[t]
\caption{The fermion fields of the standard model and their gauge quantum
numbers.}
\begin{center}\begin{tabular}[7]{cccccccc}
&&&&$\underline{SU(3)}$&$\underline{SU(2)}$&$\underline{U(1)_Y}$\\
\\
$Q_L^i=$&$\left(\begin{array}{l}u_L\\d_L\end{array}\right)$
&$\left(\begin{array}{l}c_L\\s_L\end{array}\right)$
&$\left(\begin{array}{l}t_L\\b_L\end{array}\right)$&3&2&$\frac{1}{6}$\\
\\
$u_R^i=$&$u_R$&$c_R$&$t_R$&3&1&$\frac{2}{3}$\\
\\
$d_R^i=$&$d_R$&$s_R$&$b_R$&3&1&$-\frac{1}{3}$\\
\\
$L_L^i=$&$\left(\begin{array}{l}\nu_{eL}\\e_L\end{array}\right)$
&$\left(\begin{array}{l}\nu_{\mu L}\\\mu_L\end{array}\right)$
&$\left(\begin{array}{l}\nu_{\tau
L}\\\tau_L\end{array}\right)$&1&2&$-\frac{1}{2}$\\
\\
$e_R^i=$&$e_R$&$\mu_R$&$\tau_R$&1&1&$-1$
\end{tabular}\end{center} \label{tab:sm}
\end{table}

Let's break the Lagrangian of the standard model into pieces.  First consider
the pure gauge interactions, given by
\begin{equation}
{\cal L}_{Gauge} = \frac{1}{2g_S^2}{\rm Tr}\;
G^{\mu\nu}G_{\mu\nu}+\frac{1}{2g^2}{\rm Tr}\;
W^{\mu\nu}W_{\mu\nu}-\frac{1}{4}B^{\mu\nu}B_{\mu\nu}\;, \label{Lgauge}
\end{equation}
where $G^{\mu\nu}$ is the field-strength tensor of the gluon field,
$W^{\mu\nu}$ is that of the weak-boson field, and $B^{\mu\nu}$ is that of the
hypercharge-boson field.  These terms contain the kinetic energy of the gauge
bosons and their self interactions.  Next comes the gauge interactions of the
fermion (``matter'') fields,
\begin{equation}
{\cal L}_{Matter}=i\bar Q_L^i\not\!\!DQ_L^i+i\bar u_R^i\not\!\!Du_R^i+i\bar
d_R^i\not\!\!Dd_R^i+i\bar L_L^i\not\!\!DL_L^i+i\bar e_R^i\not\!\!De_R^i\;,
\label{Lmatter}
\end{equation}
These terms contain the kinetic energy and gauge interactions of the fermions,
which depend on the fermion quantum numbers. For example,
\begin{equation}
\not\!\!DQ_L = \gamma^\mu(\partial_\mu+ig_SG_\mu+igW_\mu+i\frac{1}{6}g'
B_\mu)Q_L
\end{equation}
since the field $Q_L$ participates in all three gauge interactions. A sum on
the index $i$, which represents the generation, is implied in the Lagrangian.

We have constructed the simplest and most general Lagrangian, given the
fermion fields and gauge symmetries.\footnote{I will give a precise definition
to ``simplest'' later in this lecture.  For now, it means the minimum number of
fields and derivatives are used in each term in the Lagrangian.} The gauge
symmetries forbid masses for any of the particles. In the case of the
fermions, masses are forbidden by the fact that the left-chiral and
right-chiral components of a given fermion field have different $SU(2)\times
U(1)_Y$ quantum numbers.  For example, a mass term for the up quark,
\begin{equation}
{\cal L}=-m\bar u_Lu_R+h.c.\;,
\end{equation}
is forbidden by the fact that $u_L$ is part of the $SU(2)$ doublet $Q_L$, so
such a term violates the $SU(2)$ gauge symmetry (it also violates $U(1)_Y$).

Although we only imposed the gauge symmetry on the Lagrangian, it turns out
that it has a good deal of global symmetry as well, associated with the three
generations.  Because all fermions are massless thus far in our analysis,
there is no difference between the three generations - they are physically
indistinguishable.  This manifests itself as a global flavor symmetry of the
matter Lagrangian, Eq.~(\ref{Lmatter}), which is invariant under the
transformations
\begin{eqnarray}
&&Q_L^i \to U_{Q_L}^{ij}Q_L^j\nonumber\\
&&u_R^i \to U_{u_R}^{ij}u_R^j\nonumber\\
&&d_R^i \to U_{d_R}^{ij}d_R^j\nonumber\\
&&L_L^i \to U_{L_L}^{ij}L_L^j\nonumber\\
&&e_R^i \to U_{e_R}^{ij}e_R^j\;,\label{U(3)5}
\end{eqnarray}
where each $U$ is an arbitrary $3\times 3$ unitary matrix.
\\[7pt]
{\em Exercise 1.1 ($\ast$) Show this.}
\\[7pt]
Since there are five independent $U(3)$ symmetries, the global flavor symmetry
of the Lagrangian is $[U(3)]^5$.

The Lagrangian thus far contains only three parameters, the couplings of the
three gauge interactions.  Their approximate values (evaluated at $M_Z$) are
\begin{eqnarray*}
g_S&\approx& 1 \\
g&\approx& \frac{2}{3} \\
g'&\approx& \frac{2}{3\sqrt 3}\;.
\end{eqnarray*}
These couplings are all of order unity.

{\em Electroweak symmetry breaking} -- The theory thus far is very simple and
elegant, but it is incomplete - all particles are massless.  We now turn to
electroweak symmetry breaking, which is responsible for generating the masses
of the gauge bosons and fermions.

\begin{figure}
\begin{center}
\epsfxsize=3in \epsfbox{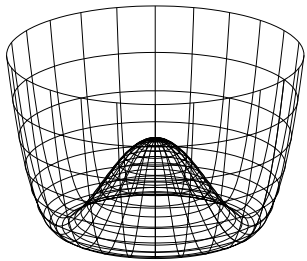}
\end{center}
\caption{The Higgs potential.  The neutral component of the Higgs field
acquires a vacuum-expectation value $\langle \phi^0\rangle = v/\sqrt 2$ on the
circle of minima in Higgs-field space.} \label{higgspotential}
\end{figure}

\begin{table}[b]
\caption{The Higgs field and its gauge quantum numbers.}
\begin{center}\begin{tabular}[7]{cccc}
&$\underline{SU(3)}$&$\underline{SU(2)}$&$\underline{U(1)_Y}$\\
\\
$\phi=\left(\begin{array}{l}\phi^+\\\phi^0\end{array}\right)$
&1&2&$\frac{1}{2}$
\end{tabular}\end{center} \label{tab:higgs}
\end{table}

In the standard model, electroweak symmetry breaking is achieved by
introducing another field into the model, the Higgs field $\phi$, with the
quantum numbers shown in Table~\ref{tab:higgs}.  The simplest and most general
Lagrangian for the Higgs field, consistent with the gauge symmetry, is
\begin{equation}
{\cal L}_{Higgs} = (D^\mu\phi)^\dagger
D_\mu\phi+\mu^2\phi^\dagger\phi-\lambda(\phi^\dagger\phi)^2\;.\label{LHiggs}
\end{equation}
The first term contains the Higgs-field kinetic energy and gauge
interactions.  The remaining terms are (the negative of) the Higgs potential,
shown in Fig.~\ref{higgspotential}.  The quadratic term in the potential has
been chosen such that the minimum of the potential lies not at zero, but on a
circle of minima
\begin{equation}
\langle\phi^0\rangle=\mu/\sqrt{2\lambda}\equiv\frac{v}{\sqrt 2} \label{vev}
\end{equation}
where $\phi^0$ is the lower (neutral) component of the Higgs doublet field.
This equation defines the parameter $v\approx 246$ GeV, the Higgs-field
vac\-uum-expectation value.  Making the substitution $\phi=(0,v/\sqrt 2)$ in
the Higgs Lagrangian, Eq.~(\ref{LHiggs}), one finds that the $W$ and $Z$ bosons
have acquired masses
\begin{equation}
M_W=\frac{1}{2}gv \hspace{2cm} M_Z=\frac{1}{2}\sqrt{g^2+g'^2}\,v \label{Wmass}
\end{equation}
from the interaction of the gauge bosons with the Higgs field.  Since we know
$g$ and $g'$, these equations determine the numerical value of $v$.

The Higgs sector of the theory, Eq.~(\ref{LHiggs}), introduces just two new
parameters, $\mu$ and $\lambda$.  Rather than $\mu$, we will use the parameter
$v$ introduced in Eq.~(\ref{vev}).  The parameter $\lambda$ is the Higgs-field
self interaction, and will not figure into our discussion.

{\em Fermion masses and mixing} -- In quantum field theory, anything that is
not forbidden is mandatory. With that in mind, there is one more set of
interactions, involving the Higgs field and the fermions. The simplest and
most general Lagrangian, consistent with the gauge symmetry, is
\begin{equation}
{\cal L}_{Yukawa} = -\Gamma_u^{ij}\bar Q_L^i\epsilon
\phi^*u_R^j-\Gamma_d^{ij}\bar Q_L^i\phi d_R^j-\Gamma_e^{ij}\bar L_L^i\phi
e_R^j + h.c. \label{LYukawa}
\end{equation}
where $\Gamma_u,\Gamma_d,\Gamma_e$ are $3\times 3$ complex matrices in
generation space.\footnote{The matrix $\epsilon = \left(\begin{array}{rr} 0 & 1 \\
-1 & 0 \end{array}\right)$ in $SU(2)$ space is needed in order for the first
term in Eq.~(\ref{LYukawa}) to respect $SU(2)$ gauge invariance.} We have
therefore apparently introduced $3\times 3\times 3\times 2=54$ new parameters
into the theory, but as we shall see, only a subset of these parameters are
physically relevant. These so-called Yukawa interactions of the Higgs field
with fermions violate almost all of the $[U(3)]^5$ global symmetry of the
fermion gauge interactions, Eq.~(\ref{Lmatter}).  The only remaining global
symmetries are the subset corresponding to baryon number
\begin{eqnarray}
&&Q_L^i \to e^{i\theta/3}Q_L^i \nonumber\\
&&u_R^i \to e^{i\theta/3}u_R^i \nonumber\\
&&d_R^i \to e^{i\theta/3}d_R^i \label{baryon}
\end{eqnarray}
and lepton number
\begin{eqnarray}
&&L_L^i \to e^{i\phi}L_L^i \nonumber\\
&&e_R^i \to e^{i\phi}e_R^i \;.\label{lepton}
\end{eqnarray}
{\em Exercise 1.2 ($\ast$) Show this.}
\\[7pt]
\indent The conservation of baryon number and lepton number follow from these
symmetries.  These symmetries are accidental; they are not put in by hand, but
rather follow automatically from the field content and gauge symmetries of the
theory.  Thus we can say that we understand why baryon number and lepton
number are conserved in the standard model.

Replacing the Higgs field with its vacuum-expectation value, $\phi=(0,v/\sqrt
2)$, in Eq.~(\ref{LYukawa}) yields
\begin{equation}
{\cal L}_{M} = -M_u^{ij}\bar u_L^iu_R^j-M_d^{ij}\bar d_L^id_R^j-M_e^{ij}\bar
e_L^ie_R^j + h.c.\;, \label{Lmass}
\end{equation}
where
\begin{equation}
M^{ij}=\Gamma^{ij}\frac{v}{\sqrt 2} \label{massmatrix}
\end{equation}
are fermion
mass matrices. The Yukawa interactions are therefore responsible for providing
the charged fermions with mass; the neutrinos, however, remain massless (we
will discuss neutrino masses shortly).

The complete Lagrangian of the standard model is the sum of the gauge, matter,
Higgs, and Yukawa interactions,
\begin{equation}
{\cal L}_{SM} = {\cal L}_{Gauge} + {\cal L}_{Matter} + {\cal L}_{Higgs} + {\cal
L}_{Yukawa}\;.
\end{equation}
This is the simplest and most general Lagrangian, given the field content and
gauge symmetries of the standard model.

Given this Lagrangian, one can proceed to calculate any physical process of
interest.  However, it is convenient to first perform field redefinitions to
make the physical content of the theory manifest.  These field redefinitions
do not change the predictions of the theory; they are analogous to a change of
variables when performing an integration.  To make the masses of the fermions
manifest, we perform unitary field redefinitions on the fields in order to
diagonalize the mass matrices in Eq.~(\ref{Lmass}):
\begin{eqnarray}
&&u_L^i = A_{u_L}^{ij}u_L'^j\;\;\;\;\;u_R^i = A_{u_R}^{ij}u_R'^j\nonumber\\
&&d_L^i = A_{d_L}^{ij}d_L'^j\;\;\;\;\;d_R^i = A_{d_R}^{ij}d_R'^j\nonumber\\
&&e_L^i = A_{e_L}^{ij}e_L'^j\;\;\;\;\;e_R^i = A_{e_R}^{ij}e_R'^j\nonumber\\
&&\nu_L^i = A_{\nu_L}^{ij}\nu_L'^j\label{fieldredef}
\end{eqnarray}
{\em Exercise 1.3 ($\ast$) Show that each matrix $A$ must be unitary in order
to preserve the form of the kinetic-energy terms in the matter Lagrangian,
Eq.~(\ref{Lmatter}), {\it e.g.}}
\begin{equation}
{\cal L}_{KE} = i\bar u_L^i\not\!\partial u_L^i\;.\label{KE}
\end{equation}
\indent Once the mass matrices are diagonalized, the masses of the fermions are
manifest.  These transformations also diagonalize the Yukawa matrices
$\Gamma$, since they are proportional to the mass matrices [see
Eq.~(\ref{massmatrix})]. However, we must consider what impact these field
redefinitions have on the rest of the Lagrangian.  They have no effect on the
pure gauge or Higgs parts of the Lagrangian, Eqs.~(\ref{Lgauge}) and
(\ref{LHiggs}), which are independent of the fermion fields. They do impact
the matter part of the Lagrangian, Eq.~(\ref{Lmatter}).  However, a subset of
these field redefinitions is the global $[U(3)]^5$ symmetry of the matter
Lagrangian; this subset therefore has no impact.

One can count how many physically-relevant parameters remain after the field
redefinitions are performed \cite{Falk:2000tx}.  Let's concentrate on the
quark sector.  The number of parameters contained in the complex matrices
$\Gamma_u,\Gamma_d$ is $2\times 3\times 3\times 2=36$. The unitary symmetries
$U_{Q_L},U_{u_R},U_{d_R}$ are a subset of the quark field redefinitions; this
subset will not affect the matter part of the Lagrangian.  There are $3\times
3\times 3$ degrees of freedom in these symmetries (a unitary $N\times N$
matrix has $N^2$ free parameters), so the total number of parameters that
remain in the full Lagrangian after field redefinitions is
\begin{equation}
2\times 3\times 3\times 2 - (3\times 3\times 3-1) = 10
\end{equation}
where I have subtracted baryon number from the subset of field redefinitions
that are symmetries of the matter Lagrangian.  Baryon number is a symmetry of
the Yukawa Lagrangian, Eq.~(\ref{LYukawa}), and hence cannot be used to
diagonalize the mass matrices.
\\[7pt]
{\em Exercise 1.4 ($\ast$) Show that the quark field redefinitions {\it are}
the symmetries $U_{Q_L},U_{u_R},U_{d_R}$ if $A_{u_L}=A_{d_L}$.}
\\[7pt]
\indent The ten remaining parameters correspond to the six quark masses and the
four parameters of the Cabibbo-Kobayashi-Maskawa (CKM) matrix (three mixing
angles and one $CP$-violating phase).  The CKM matrix is $V\equiv
A_{d_L}^\dagger A_{u_L}$; we see that this matrix is unity if
$A_{u_L}=A_{d_L}$, as expected from Exercise 1.4.
\\[7pt]
{\em Exercise 1.5 ($\ast$) Show that $V$ is unitary.}
\\[7pt]
\indent The mass matrices are related to the Yukawa matrices by
Eq.~(\ref{massmatrix}).  If we make the natural assumption that the Yukawa
matrices contain elements of order unity (like the gauge couplings), we expect
the fermion masses to be of ${\cal O}(v)$, just like $M_W$ and $M_Z$ [see
Eq.~(\ref{Wmass})]. This is not the case; only the top quark has such a large
mass.  We see that, from the point of view of the standard model, the question
is not why the top quark is so heavy, but rather why the other fermions are so
light.

Similarly, for a generic Yukawa matrix, one expects the field redefinitions
that diagonalize the mass matrices to yield a CKM matrix with large mixing
angles.  Again, this is not the case; the measured angles are
\cite{Hagiwara:pw}
\begin{eqnarray*}
\theta_{12}&\approx& 13^\circ\\
\theta_{23}&\approx& 2.3^\circ\\
\theta_{13}&\approx& 0.23^\circ\\
\delta&\approx& 60^\circ
\end{eqnarray*}
which, with the exception of the $CP$-violating phase $\delta$, are
small.\footnote{The phase $\delta$ is the same as the angle $\gamma$ of the
so-called unitarity triangle.} The question is not why these angles are
nonzero, but rather why they are so small.

The fermion masses and mixing angles strongly suggest that there is a deeper
structure underlying the Yukawa sector of the standard model.  Surely there is
some explanation of the peculiar pattern of fermion masses and mixing angles.
Since the standard model can accommodate any masses and mixing angles, we must
seek an explanation from physics beyond the standard model.

{\em Beyond the Standard Model} -- Let us back up and ask: why did we stick to
the simplest terms in the Lagrangian?  The obsolete answer is that these are
the renormalizable terms. Renormalizability is a stronger constraint than is
really necessary. The modern answer, which is much simpler, is dimensional
analysis \cite{Weinberg:pi}.

We'll work with units such that $\hbar=c=1$.
\\[7pt]
{\em Exercise 1.6 ($\ast$) - Show that length has units of mass$^{-1}$, and
hence $\partial_\mu = \partial/\partial x^\mu$ has units of mass.}
\\[7pt]
Since the action has units of $\hbar = 1$, the Lagrangian must have units of
${\rm mass}^4$, since
\begin{equation}
S=\int d^4x\;{\cal L}\;.
\end{equation}
From the kinetic energy terms in the Lagrangian for a generic scalar ($\phi$),
fermion ($\psi$), and gauge boson ($A^\mu$),
\begin{equation}
{\cal L}_{KE} = \partial^\mu\phi^*\partial_\mu\phi +
i\bar\psi\not\!\partial\psi - \frac{1}{2}(\partial^\mu A^\nu\partial_\mu A_\nu
+ \partial^\mu A^\nu\partial_\nu A_\mu)
\end{equation}
we can deduce the dimensionality of the various fields:
\begin{eqnarray*}
&&{\rm dim}\;\phi = {\rm mass} \\
&&{\rm dim}\;\psi = {\rm mass}^{3/2} \\
&&{\rm dim}\; A^\mu = {\rm mass}\;.
\end{eqnarray*}
All operators (products of fields) in the Lagrangian of the Standard Model are
of dimension four, except the operator $\phi^\dagger\phi$ in the Higgs
potential, which is of dimension two.  The coefficient of this term, $\mu^2$,
is the only dimensionful parameter in the standard model; it (or, equivalently,
$v\equiv\mu/\sqrt\lambda$) sets the scale of all particle masses.

Imagine that the Lagrangian at the weak scale is an expansion in some large
mass scale $M$,
\begin{equation}
{\cal L}={\cal L}_{SM}+\frac{1}{M}{\rm dim}\; 5+\frac{1}{M^2}{\rm dim}\;
6+\cdots\;, \label{Lexpansion}
\end{equation}
where ${\rm dim}\; n$ represents all operators of dimension $n$.  By
dimensional analysis, the coefficient of an operator of dimension $n$ has
dimension ${\rm mass}^{4-n}$, since the Lagrangian has dimension ${\rm
mass}^4$.  At energies much less than $M$, the dominant terms in this
Lagrangian will be those of ${\cal L}_{SM}$; the other terms are suppressed by
an inverse power of $M$. This is the modern reason why we believe the
``simplest'' terms in the Lagrangian are the dominant ones.

The least suppressed terms in the Lagrangian beyond the standard model are of
dimension five.  We should therefore expect our first observation of physics
beyond the standard model to come from these terms.  Given the field content
and gauge symmetries of the standard model, there is only one such term:
\begin{equation}
{\cal L}_5=\frac{c^{ij}}{M}L_L^{iT}\epsilon\phi C \phi^T\epsilon L_L^j+h.c.\;,
\label{L5}
\end{equation}
where $c^{ij}$ is a dimensionless matrix in generation space.\footnote{The
$2\times 2$ matrix $\epsilon$ in $SU(2)$ space was introduced in an earlier
footnote. The $4\times 4$ matrix $C=\left(\begin{array}{rr} -\epsilon & 0 \\
0 & \epsilon \end{array}\right)$ in Dirac space is needed for Lorentz
invariance.}
\\[7pt]
{\em Exercise 1.7 ($\ast\ast$) - Show that a similar term, with $L_L$ replaced
by $Q_L$, is forbidden by $SU(3)\times U(1)_Y$ gauge symmetry.}
\\[7pt]
This dimension-five operator contains the Higgs-doublet field twice and the
lepton-doublet field twice.
\\[7pt]
{\em Exercise 1.8 ($\ast$) - Show that ${\cal L}_5$ violates lepton number.}
\\[7pt]
Replacing the Higgs-doublet field with its vacuum-expectation value, $\phi =
(0,v/\sqrt 2)$, yields
\begin{equation}
{\cal L}_5=-\frac{c^{ij}}{2}\frac{v^2}{M}\nu_L^{iT} C \nu_L^j+h.c.\;.
\end{equation}
This is a Majorana mass term for the neutrinos.  The recent observation of
neutrino oscillations, which requires nonzero neutrino mass, is indeed our
first observation of physics beyond the standard model.
\\[7pt]
{\em Exercise 1.9 ($\ast\ast\ast$) Show that the Maki-Nakagawa-Sakata (MNS)
matrix (the analogue of the CKM matrix in the lepton sector) has 6
physically-relevant parameters. (Note: $c^{ij}$ is a complex, symmetric
matrix.)}
\\[7pt]
\indent The moral is that when we are searching for deviations from the
standard model, what we are really doing is looking for the effects of
higher-dimension operators.  Although there is only one operator of dimension
five, there are dozens of operators of dimension six, some of which are listed
below \cite{Buchmuller:1985jz}:
\begin{eqnarray*}
&&\bar L^i\gamma^\mu L^j\bar L^k\gamma_\mu L^m\\
&&\bar L^i\gamma^\mu L^j\bar Q^k\gamma_\mu Q^m\\
&&i\bar Q^i\gamma_\mu D_\nu G^{\mu\nu}Q^j\\
&&{\rm Tr}\;G^{\mu\nu}G_{\nu\rho}G^\rho_\mu\\
&&\phi^\dagger\phi {\rm Tr}\;W^{\mu\nu}W_{\mu\nu}\\
&&\phi^\dagger D_\mu\phi \bar e_R^i\gamma^\mu e_R^j\;.
\end{eqnarray*}
Thus far, none of the effects of any of these operators have been observed.
The best we can do is set lower bounds on $M$ (assuming some dimensionless
coefficient). These lower bounds range from 1 TeV to $10^{16}$ GeV, depending
on the operator.  As we explore nature at higher energy and with higher
accuracy, we hope to begin to see the effects of some of these dimension-six
operators.

The mass scale $M$ corresponds to the mass of a particle that is too heavy to
observe directly.  At energies greater than $M$, the expansion of
Eq.~(\ref{Lexpansion}) is no longer useful, as each successive term is larger
than the previous.  Instead, one must explicitly add the new field of mass $M$
to the model. For example, if nature is supersymmetric at the weak scale, one
must add the superpartners of the standard-model fields to the theory and
include their interactions in the Lagrangian.  If we raise the mass scale of
the superpartners to be much greater than the weak scale, then we can no longer
directly observe the superpartners, and we return to a description in terms of
standard-model fields, with an expansion of the Lagrangian in inverse powers
of the mass scale of the superpartners, $M$.

\section{Virtual Top Quark}\label{sec:virtual}

The top quark plays an important role in precision electroweak analyses.  In
this lecture I hope to clarify this sometimes confusing subject.

Recall from the previous lecture that the gauge, matter, and Higgs sectors of
the standard model depend on only five parameters: the three gauge couplings,
$g_S$, $g$, $g'$, and the Higgs-field vacuum-expectation value and self
interaction, $v$ and $\lambda$.  At tree level, all electroweak quantities
depend on just three of these parameters, $g$, $g'$, and $v$.  We use the
three best-measured electroweak quantities to determine these three parameters
at tree level:
\begin{eqnarray*}
\alpha&=&\frac{1}{4\pi}\frac{g^2g'^2}{g^2+g'^2}=\frac{1}{137.03599976(50)}\\
G_F&=&\frac{1}{\sqrt 2v^2}=1.16637(1)\times 10^{-5}\;{\rm GeV}^{-2}\\
M_Z&=&\frac{1}{2}\sqrt{g^2+g'^2}\, v=91.1876(21)\;{\rm GeV}\;,
\end{eqnarray*}
where the uncertainty is given in parentheses.  The value of $\alpha$ is
extracted from low-energy experiments, $G_F$ is extracted from the muon
lifetime, and $M_Z$ is measured from $e^+e^-$ annihilation near the $Z$ mass.
From these three quantities, we can predict all other electroweak quantities
at tree level. For example, the $W$ mass is
\begin{equation}
M_W^2=\frac{1}{4}g^2v^2=\frac{1}{2}M_Z^2\left(1+\sqrt{1-\frac{4\pi\alpha}{\sqrt
2G_FM_Z^2}}\right)\;.\label{MW}
\end{equation}
{\em Exercise 2.1 ($\ast$) Verify the expression for $M_W$ in terms of
$\alpha$, $G_F$, and $M_Z$.}
\\[7pt]
\indent A more civilized expression for $M_W$ is obtained by {\em defining}
\begin{equation}
s_W^2\equiv 1-\frac{M_W^2}{M_Z^2}\;.\label{s2w}
\end{equation}
This is the so-called ``on-shell'' definition\footnote{So called because it is
defined in terms of physical, or ``on shell,'' quantities.}  of
$\sin^2\theta_W$; it has a numerical value of $s_W^2=0.2228(4)$. Using this
parameter, we can write a simpler expression than Eq.~(\ref{MW}) for $M_W$ at
tree level:
\begin{equation}
M_W^2=\frac{\frac{\pi\alpha}{\sqrt 2G_F}}{s_W^2}\;.\label{MW2}
\end{equation}
{\em Exercise 2.2 ($\ast$) - Verify this equation.}
\\[7pt]
\begin{figure}[b]
\begin{center}
\epsfxsize=3.5in \epsfbox{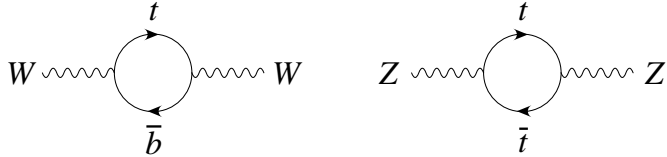}
\end{center}
\caption{Virtual top-quark loops contribute to the $W$ and $Z$ masses.}
\label{toploops}
\end{figure}
\indent At one loop this expression is modified:
\begin{equation}
M_W^2=\frac{\frac{\pi\alpha}{\sqrt 2G_F}}{s_W^2(1-\Delta r)}\;,\label{MWloop}
\end{equation}
where $\Delta r$ contains the one-loop corrections.  The top quark makes a
contribution to $\Delta r$ via the one-loop diagrams shown in
Fig.~\ref{toploops}, which contribute to the $W$ and $Z$ masses:
\begin{equation}
(\Delta r)_{\rm top}\approx -\frac{3G_Fm_t^2}{8\sqrt
2\pi^2}\frac{1}{t_W^2}\;,\label{deltartop}
\end{equation}
where $t_W^2\equiv \tan^2\theta_W$.  This one-loop correction depends
quadratically on the top-quark mass.

\begin{figure}[t]
\begin{center}
\epsfxsize=3.5in \epsfbox{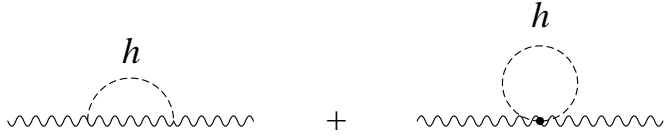}
\end{center}
\caption{Virtual Higgs-boson loops contribute to the $W$ and $Z$ masses.}
\label{higgsloops}
\end{figure}

The Higgs boson also contributes to $\Delta r$ via the one-loop diagrams in
Fig.~\ref{higgsloops}:
\begin{equation}
(\Delta r)_{\rm Higgs}\approx \frac{11G_FM_Z^2c^2_W}{24\sqrt
2\pi^2}\ln\frac{m_h^2}{M_Z^2}\;,
\end{equation}
where $c_W^2\equiv \cos^2\theta_W$.  This one-loop correction depends only
logarithmically on the Higgs-boson mass, so $\Delta r$ is not nearly as
sensitive to $m_h$ as it is to $m_t$.

Due to the contributions of the top quark and the Higgs boson to $\Delta r$,
in order to predict $M_W$ at one loop via Eq.~(\ref{MWloop}) we need not just
$\alpha$, $G_F$, $M_Z$, but also $m_t$ and $m_h$.  Turning this around, in
order to predict $m_h$, we need $\alpha$, $G_F$, $M_Z$, and $m_t$, $M_W$. Thus
a precision measurement of $m_t$ and $M_W$ can be used to predict the Higgs
mass.

\begin{figure}[t]
\begin{center}
\epsfxsize=3in \epsfbox{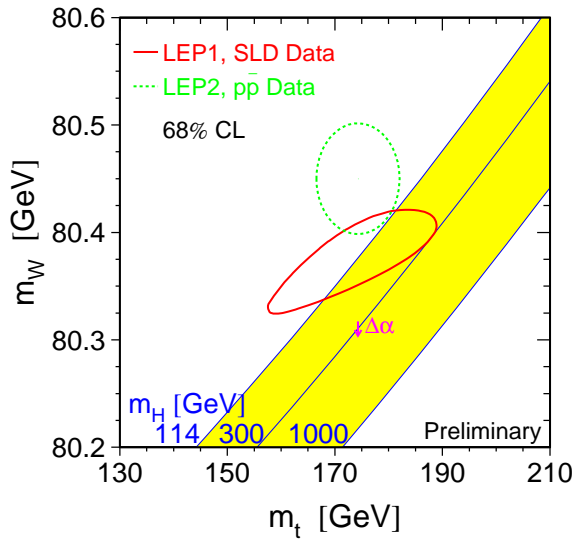}
\end{center}
\caption{Lines of constant Higgs mass on a plot of $M_W$ vs.~$m_t$.  The
dashed ellipse is the $68\%$ CL direct measurement of $M_W$ and $m_t$.  The
solid ellipse is the $68\%$ CL indirect measurement from precision electroweak
data.  From http://lepewwg.web.cern.ch/LEPEWWG/.} \label{mwmt}
\end{figure}

I show in Fig.~\ref{mwmt} a plot of $M_W$ {\it vs.} $m_t$, indicating lines of
constant Higgs mass.\footnote{The small arrow labeled $\Delta\alpha$ in that
plot indicates the uncertainty in the lines of constant Higgs mass due to the
uncertainty in $\alpha(M_Z)$.} The dashed ellipse indicates the $68\%$ CL
measurements of $M_W$ and $m_t$,
\begin{eqnarray*}
M_W&=&80.451(33)\;{\rm GeV}\\
m_t&=&174.3(5.1)\;{\rm GeV}
\end{eqnarray*}
(I will return to the solid ellipse momentarily).  As you can see, the direct
measurements of $M_W$ and $m_t$ favor a light Higgs boson.
\\[7pt]
{\em Exercise 2.3 ($\ast\ast\ast$) - Derive the slope of the lines of constant
Higgs mass on the plot of $M_W$ {\it vs.} $m_t$.  Evaluate it numerically and
compare with the plot.  [Hint: Derive $dM_W^2/dm_t^2$ and then evaluate
$dM_W/dm_t=(m_t/M_W)dM_W^2/dm_t^2$ numerically.  Be careful to use
Eq.~(\ref{s2w}) for $s_W^2$ in Eq.~(\ref{MWloop}) (you can neglect the
dependence of $t_W^2$ on $M_W$ in Eq.~(\ref{deltartop}) since $(\Delta r)_{\rm
top}$ is a small correction).]}
\\[7pt]
\indent {\em Neutral current} -- Rather than using the direct measurements of
$M_W$ and $m_t$ to infer the Higgs-boson mass, one can use other electroweak
quantities.  The Fermi constant, $G_F$, is extracted from muon decay, which is
a charged-current weak interaction.  That leaves the neutral-current weak
interaction as another quantity of interest. There is an enormous wealth of
data on neutral-current weak interactions, such as $e^+e^-$ annihilation near
the $Z$ mass, $\nu N$ and $eN$ deep-inelastic scattering, $\nu e$ elastic
scattering, atomic parity violation, and so on \cite{Hagiwara:pw}.

\begin{figure}
\begin{center}
\begin{picture}(220,110)
\put(0,0){\epsfxsize=1.3in \epsfbox{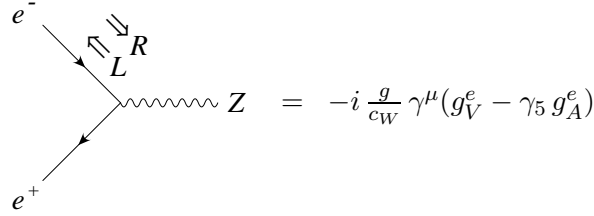}} \put(105,55){$= \;\; -i
\, \frac{g}{c_W} \, \gamma^{\mu}
                         (g_V^e - \gamma_5 \, g_A^e)$}
\end{picture}
\end{center}
\caption{Neutral-current coupling of an electron to the $Z$ boson.  A
left-handed electron has negative helicity, a right-handed electron has
positive helicity.} \label{LRasymmetry}
\end{figure}

Let's consider a simple and very relevant example, the left-right asymmetry in
$e^+e^-$ annihilation near the $Z$ mass, shown in Fig.~\ref{LRasymmetry}.  Left
and right refer to the helicity of the incident electron, either negative
(left) or positive (right).  The asymmetry is defined in terms of the total
cross section for a negative-helicity or positive-helicity electron to
annihilate with an unpolarized positron and produce a $Z$ boson,
\begin{eqnarray}
A_{LR}&\equiv & \frac{\sigma_L-\sigma_R}{\sigma_L+\sigma_R}\nonumber\\
&=&\frac{2g_V^eg_A^e}{g_V^{e2}+g_A^{e2}}\;,
\end{eqnarray}
where
\begin{eqnarray}
&&g_V^e=\sqrt{\rho_e}\left(-\frac{1}{2}+2\kappa_e s_W^2\right)\nonumber\\
&&g_A^e=\sqrt{\rho_e}\left(-\frac{1}{2}\right)
\end{eqnarray}
are the vector and axial-vector couplings of the electron to the $Z$ boson. At
tree level, $\rho_e=\kappa_e=1$, but there are one-loop corrections.  The
correction quadratic in the top-quark mass is
\begin{eqnarray}
&&(\rho_e)_{\rm top}\approx 1+\frac{3G_Fm_t^2}{8\sqrt 2\pi^2} \nonumber \\
&&(\kappa_e)_{\rm top}\approx 1+\frac{3G_Fm_t^2}{8\sqrt
2\pi^2}\frac{1}{t^2_W}\;.
\end{eqnarray}

Different neutral-current measurements have different dependencies on $m_t$ and
$m_h$, so by combining two or more measurements one can extract both $m_t$ and
$m_h$.  The solid ellipse in Fig.~\ref{mwmt} represents the $68\%$ CL
constraint from all neutral-current measurements combined.  It is in good
agreement with the direct measurements of $M_W$ and $m_t$, and strengthens the
case for a light Higgs boson.  Combining all precision electroweak data, one
finds 45 GeV $\le m_h\le 191$ GeV \cite{Hagiwara:pw}.

Historically, neutral-current data were used to successfully predict the
top-quark mass several years before it was discovered.  This is a good reason
to trust the prediction of a light Higgs boson from precision electroweak
analyses.

It is also significant that the two ellipses in Fig.~\ref{mwmt} lie on or near
the lines of constant Higgs mass (within the allowed range of the Higgs mass).
These measurements could have ended up far from those lines, thereby
disproving the existence of the hypothetical Higgs boson. Instead, these
measurements bolster our belief in the standard model in general, and in the
Higgs boson in particular.

{\em $\overline{\rm MS}$ scheme} -- Before we leave this topic, let's discuss
the other most often-used definition of $\sin^2\theta_W$.  This is the
minimal-subtraction-bar ($\overline{\rm MS}$) scheme, so-called due to the
simple way in which ultraviolet divergences in loop diagrams are subtracted.
\\[7pt]
{\em Exercise 2.4 ($\ast$) - Show that}
\begin{equation}
\sin^2\theta_W=\frac{g'^2}{g^2+g'^2}
\end{equation}
{\em at tree level.}
\\[7pt]
The $\overline{\rm MS}$ scheme promotes this to the
definition of $\sin^2\theta_W$:
\begin{equation}
\hat s^2_Z\equiv \frac{g'^2(M_Z)}{g^2(M_Z)+g'^2(M_Z)}
\end{equation}
where the gauge couplings are evaluated at the $Z$ mass.  Its numerical value
is $\hat s^2_Z=0.23113(15)$.

The analogues of Eqs.~(\ref{MWloop}) and (\ref{s2w}) in the $\overline{\rm MS}$
scheme are
\begin{eqnarray}
M_W^2&=&\frac{\frac{\pi\alpha}{\sqrt 2G_F}}{\hat s^2_Z(1-\Delta\hat r_W)}\label{MSbar}\\
M_Z^2&=&\frac{M_W^2}{\hat c_Z^2\hat\rho}\;.\label{rho}
\end{eqnarray}
Unlike its on-shell analogue $\Delta r$, the one-loop quantity $\Delta\hat r_W$
has no quadratic dependence on the top-quark mass.  This appears instead in
the quantity $\hat\rho$ (which is unity in the on-shell scheme):
\begin{equation}
\hat\rho\approx 1+\frac{3G_Fm_t^2}{8\sqrt 2\pi^2}\;.
\end{equation}
Although the quadratic dependence on the top-quark mass has been shifted from
one relation to another, the physical predictions, such as the constraint on
the Higgs mass, remain unchanged.
\\[7pt]
{\em Exercise 2.5 ($\ast\ast\ast$) - Repeat Exercise 2.3 in the $\overline{\rm
MS}$ scheme. [Note that $\hat s^2_Z$ depends on $M_W$ via Eq.~(\ref{rho}).]}

\section{Top Strong Interactions}\label{sec:strong}

We now begin to discuss the study of the top quark itself.  In the
introduction we listed several reasons why the top quark is an interesting
object to study.  The strategy that follows from these motivations is to get to
know the top quark by measuring everything we can about it, and comparing with
the predictions of the standard model.  This program will occupy a large
portion of our efforts at the Fermilab Tevatron and the CERN Large Hadron
Collider (LHC).  In this section I discuss some of the measurements that can
be made at these machines related to the strong interactions of the top quark,
and in the next section I turn to its weak interactions.

\begin{figure}[b]
\begin{center}
\epsfxsize=4.7in \epsfbox{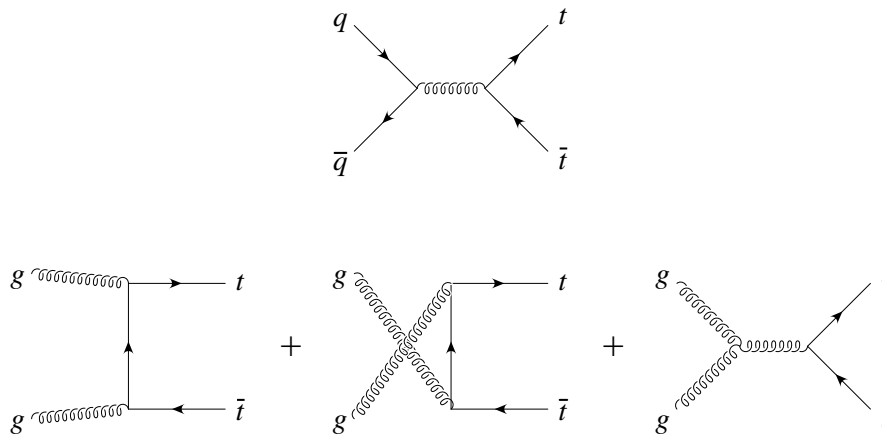}
\end{center}
\caption{Top-quark production via the strong interaction at hadron colliders
proceeds through quark-antiquark annihilation (upper diagram) and gluon fusion
(lower diagrams).} \label{topproduction}
\end{figure}

The top quark is produced at hadron colliders primarily via the strong
interaction. The Feynman diagrams for the two contributing subprocesses,
quark-antiquark annihilation and gluon fusion, are shown in
Fig.~\ref{topproduction}. In Table~\ref{tab:topsigma} I give the predicted
cross sections, at next-to-leading-order (NLO) in QCD, for $m_t=175$ GeV.  I
also show the percentage of the cross section that results from each of the two
subprocesses.  At the Tevatron, the quark-antiquark-annihilation subprocess
dominates; at the LHC, gluon fusion reigns.  To understand why this is, we
need to discuss the parton model of the proton.

\begin{table}[t]
\caption{Cross sections, at next-to-leading-order in QCD, for top-quark
production via the strong interaction at the Tevatron and the LHC
\cite{Bonciani:1998vc}.  Also shown is the percentage of the total cross
section from the quark-antiquark-annihilation and gluon-fusion subprocesses.}
\begin{center}\begin{tabular}[7]{c|c|c|c}
\hline\hline
&$\sigma_{\rm NLO}$ (pb)&$q\bar q\to t\bar t$&$gg\to t\bar t$\\
\hline
Tevatron ($\sqrt s=1.8$ TeV $p\bar p$)&$4.87\pm 10\%$&$90\%$&$10\%$ \\
Tevatron ($\sqrt s=2.0$ TeV $p\bar p$)&$6.70\pm 10\%$&$85\%$&$15\%$ \\
LHC ($\sqrt s=14$ TeV $pp$)&$803\pm 15\%$&$10\%$&$90\%$\\ \hline\hline
\end{tabular}\end{center} \label{tab:topsigma}
\end{table}

\begin{figure}[b]
\begin{center}
\epsfxsize=2in \epsfbox{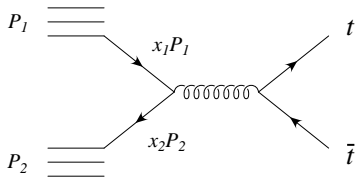}
\end{center}
\caption{The parton-model description of top-quark pair production.  A quark
carrying fraction $x_1$ of the proton's momentum $P_1$ annihilates with an
antiquark carrying fraction $x_2$ of the antiproton's momentum $P_2$.}
\label{partonmodel}
\end{figure}

The parton model is shown schematically in Fig.~\ref{partonmodel}, where I
illustrate how a proton-antiproton collision results in a $t\bar t$ pair
produced via the quark-antiquark-annihilation subprocess.  The proton is
regarded as a collection of quarks, antiquarks, and gluons (collectively
called partons), each carrying some fraction $x$ of the proton's
four-momentum.  Figure~\ref{partonmodel} shows a proton of four-momentum $P_1$
colliding with an antiproton of four-momentum $P_2$.
\\[7pt]
{\em Exercise 3.1 ($\ast$) - Show that}
\begin{equation}
S\equiv (P_1+P_2)^2\approx 2P_1\cdot P_2 \label{S}
\end{equation}
{\em (neglecting the proton mass) is the square of the total energy in the
center-of-momentum frame.}
\\[7pt]
The quark is carrying fraction $x_1$ of the proton's four-momentum, the
antiquark fraction $x_2$ of the antiproton's four-momentum. The square of the
total energy of the partonic subprocess (in the partonic center-of-momentum
frame) is similarly
\begin{equation}
\hat s\equiv (x_1P_1+x_2P_2)^2\approx 2x_1x_2P_1\cdot
P_2=x_1x_2S\;.\label{shat}
\end{equation}
Since there has to be at least enough energy to produce a $t\bar t$ pair at
rest, $\hat s\ge 4m_t^2$.  It follows from Eq.~(\ref{shat}) that
\begin{equation}
x_1x_2 =\frac{\hat s}{S}\ge \frac{4m_t^2}{S}\;.\label{x1x2}
\end{equation}
Since the probability of finding a quark of momentum-fraction $x$ in the
proton falls off with increasing $x$, the typical value of $x_1x_2$ is near the
threshold for $t\bar t$ production.  Setting $x_1\approx x_2 =x$ in
Eq.~(\ref{x1x2}) gives
\begin{equation}
x\approx \frac{2m_t}{\sqrt S}\label{typicalx}
\end{equation}
as the typical value of $x$ for $t\bar t$ production.

\begin{figure}[t]
\begin{center}
\input{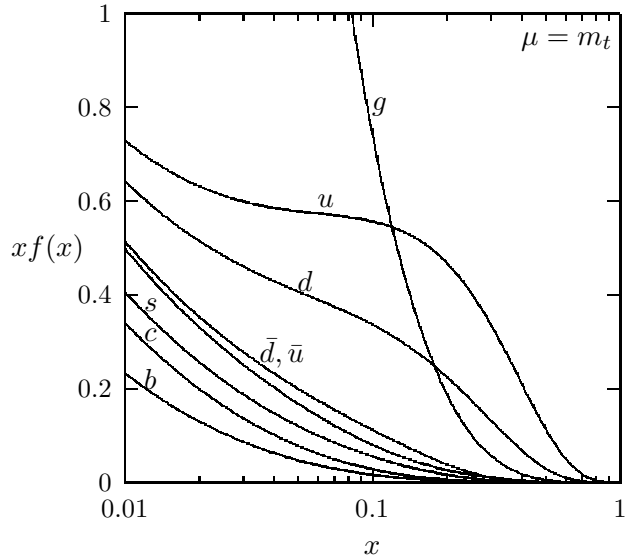}
\end{center}
\caption{Parton distribution functions at the scale $\mu=m_t$, relevant for
top-quark production.} \label{pdfs}
\end{figure}

Figure~\ref{pdfs} shows the parton distribution functions in the proton for all
the different species of partons.\footnote{I will explain in
Section~\ref{sec:weak} the presence of antiquarks in the proton, as well as
strange, charm, and bottom quarks.}  The probability of finding a given parton
species with momentum fraction between $x$ and $x+dx$ is $f(x)dx$.  [What is
plotted in Fig.~\ref{pdfs} is actually $xf(x)$].  The parton distribution
functions also depend on the relevant scale of the process, $\mu$, which for
top-quark production is of order $m_t$.

The typical value of $x$ for top-quark production may be computed from
Eq.~(\ref{typicalx}).  For the typical value of $x$ at the Tevatron, $x\approx
0.18$, the up distribution function is larger than that of the gluon, and the
down distribution function is comparable to it.  This explains why
quark-antiquark annihilation dominates at the Tevatron.  In contrast, for the
typical value of $x$ at the LHC, $x\approx 0.025$, the gluon distribution
function is much larger than those of the quarks; this explains why gluon
fusion reigns at the LHC.

\begin{figure}[b]
\begin{center}
\begin{picture}(220,110)
\put(0,0){\epsfxsize=1.3in \epsfbox{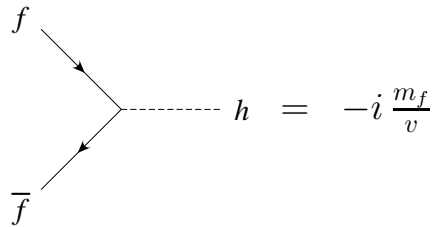}} \put(105,47.5){\Large $=
\;\; -i \, \frac{m_f}{v}$}
\end{picture}
\end{center}
\caption{The coupling of the Higgs boson to fermions.} \label{higgscoupling}
\end{figure}

\begin{figure}[t]
\begin{center}
\epsfxsize=2in \epsfbox{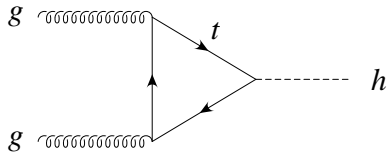}
\end{center}
\caption{Higgs-boson production via gluon fusion through a top-quark loop.}
\label{gghiggs}
\end{figure}

{\em Higgs and top} -- I mentioned in the introduction that the top quark
could be used to discover the Higgs boson.  To derive the coupling of the
Higgs boson to fermions, write the Higgs-doublet field as
\begin{equation}
\phi=\left(\begin{array}{c}0\\\frac{1}{\sqrt
2}(v+h)\end{array}\right)\label{higgsfield}
\end{equation}
where $h$ is the Higgs boson, which corresponds to oscillations about the
vacuum-expectation value of the field, Eq.~(\ref{vev}).  Inserting this
expression for $\phi$ into the Yukawa Lagrangian, Eq.~(\ref{LYukawa}), yields
the desired coupling, shown in Fig.~\ref{higgscoupling}.
\\[7pt]
{\em Exercise 3.2 ($\ast\ast\ast$) - Show that the coupling of the Higgs boson
to fermions is as given in Fig.~\ref{higgscoupling}. [Hint: Recall
$M^{ij}=\Gamma^{ij}v/\sqrt 2$, Eq.~(\ref{massmatrix}).]}
\\[7pt]
\indent The Feynman diagrams for Higgs-boson production in association with a
$t\bar t$ pair are the same as those of Fig.~\ref{topproduction}, but with a
Higgs boson attached to the top quark or antiquark.   The Higgs boson can also
be produced by itself via its coupling to a virtual top-quark loop, as shown in
Fig.~\ref{gghiggs}. Remarkably, this is the largest source of Higgs bosons at
the Tevatron or the LHC.  It is amusing that the virtual top quark points to
the existence of a light Higgs boson, as discussed in the previous section, and
may also help us discover the Higgs boson.

\begin{figure}[b]
\begin{center}
\epsfxsize=1.8in \epsfbox{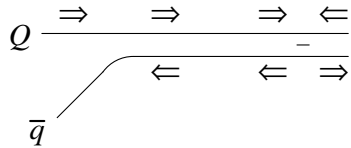}
\end{center}
\caption{A heavy quark hadronizes with a light quark of the opposite spin,
then evolves into a spin-zero meson.} \label{hadronization}
\end{figure}

{\em Top-quark spin} -- One of the remarkable features of the top quark is that
it is the only quark whose spin is directly observable.  This is a consequence
of its very short lifetime, $\Gamma_t^{-1}\approx (1.5\; {\rm GeV})^{-1}$.
Figure~\ref{hadronization} shows an example of the evolution of a heavy quark
of a definite spin after it is produced in a hard-scattering collision. On a
time scale of order $\Lambda_{QCD}^{-1}\approx (200\, {\rm MeV})^{-1}$, the
heavy quark picks up a light antiquark of the opposite spin from the vacuum and
hadronizes into a meson. Some time later, on the order of
$(\Lambda_{QCD}^2/m_Q)^{-1}\approx (1 \;{\rm MeV})^{-1}$ (for $m_Q=m_t$), the
spin-spin interaction between the heavy quark and the light
quark\footnote{This is the QCD analogue of the spin-spin interaction that
produces the hyperfine splitting in atomic physics.} cause the meson to evolve
into a spin-zero state, $(\mid\uparrow\downarrow\rangle -
\mid\downarrow\uparrow\rangle)/\sqrt 2$, thereby depolarizing the heavy quark
\cite{Falk:1993rf}. The top quark is the only quark that decays before it has a
chance to depolarize (or even hadronize), so its spin is observable in the
angular distribution of its decay products.\footnote{Actually, the spin of a
long-lived heavy quark is observable if it hadronizes into a baryon, such as a
$\Lambda_b$.}

Let's discuss the spin of a fermion in some detail.  For a moving fermion, it
is conventional to use the helicity basis, in which the spin quantization axis
is the direction of motion of the fermion.  The free fermion field may be
decomposed into states of definite four-momentum,
\begin{equation}
\psi(x) = \int\frac{d^3p}{(2\pi)^3\sqrt{2E}}\sum_{\lambda=\pm}(a_p^\lambda
u_\lambda (p)e^{-ip\cdot x}+b_p^{\lambda\dagger}v_\lambda (p)e^{ip\cdot x})\;,
\end{equation}
where the sum is over positive and negative helicity, $a_p^\lambda$ and
$b_p^{\lambda\dagger}$ are the annihilation and creation operators for a
fermion and an antifermion, and $u_\lambda (p)$ and $v_\lambda (p)$ are the
momentum-space spinors for a fermion and an antifermion.  These spinors are
given explicitly in Table~\ref{tab:spinors}, in the representation where the
Dirac matrices are \cite{Peskin}
\begin{equation}
\gamma^0=\left(\begin{array}{cc} 0 & 1 \\
1 & 0 \end{array}\right) \;\;\;
\gamma^i=\left(\begin{array}{cc} 0 & \sigma^i \\
-\sigma^i & 0 \end{array}\right) \;\;\;
\gamma_5=\left(\begin{array}{cc} -1 & 0 \\
0 & 1 \end{array}\right) \;\;\;,
\end{equation}
where each entry in the above matrices is itself a $2\times 2$ matrix.

\begin{table}[b]
\caption{Spinors for a fermion of energy $E$ and three-momentum of magnitude
${\rm p}$ pointing in the $(\theta,\phi)$ direction.  The spinors $u_\lambda
(p)$ and $v_\lambda (p)$ correspond to fermions and antifermions of helicity
$\lambda{1\over 2}$.}
\begin{center}\begin{tabular}[7]{cccc}
\\
$u_+(p)=$&$\left(\begin{array}{c}\sqrt{E-{\rm p}}
\left(\begin{array}{c}\cos{\theta\over 2}
\\e^{i\phi}\sin{\theta\over 2}\end{array}\right)\\
\sqrt{E+{\rm p}}\left(\begin{array}{c}\cos{\theta\over 2}
\\e^{i\phi}\sin{\theta\over 2}\end{array}\right)
\end{array}\right)$
& $u_-(p)=$&$\left(\begin{array}{c}\sqrt{E+{\rm p}}
\left(\begin{array}{c}\sin{\theta\over 2}
\\-e^{i\phi}\cos{\theta\over 2}\end{array}\right)\\
\sqrt{E-{\rm p}}\left(\begin{array}{c}\sin{\theta\over 2}
\\-e^{i\phi}\cos{\theta\over 2}\end{array}\right)
\end{array}\right)$
\\ \\
$v_+(p)=$&$\left(\begin{array}{c}\sqrt{E+{\rm p}}
\left(\begin{array}{c}-e^{-i\phi}\sin{\theta\over 2}
\\\cos{\theta\over 2}\end{array}\right)\\
-\sqrt{E-{\rm p}}\left(\begin{array}{c}-e^{-i\phi}\sin{\theta\over 2}
\\\cos{\theta\over 2}\end{array}\right)
\end{array}\right)$
& $v_-(p)=$&$\left(\begin{array}{c}\sqrt{E-{\rm p}}
\left(\begin{array}{c}e^{-i\phi}\cos{\theta\over 2}
\\\sin{\theta\over 2}\end{array}\right)\\
-\sqrt{E+{\rm p}}\left(\begin{array}{c}e^{-i\phi}\cos{\theta\over 2}
\\\sin{\theta\over 2}\end{array}\right)
\end{array}\right)$
\\
\end{tabular}\end{center} \label{tab:spinors}
\end{table}

We used the concept of chirality when formulating the standard model in
Section~\ref{sec:sm}.  In the representation of the Dirac matrices given above,
\begin{eqnarray}
\psi_L\equiv \frac{1-\gamma_5}{2}\psi= \left(\begin{array}{cc} 1 & 0 \\
0 & 0 \end{array}\right)\psi\\
\psi_R\equiv \frac{1+\gamma_5}{2}\psi= \left(\begin{array}{cc} 0 & 0 \\
0 & 1 \end{array}\right)\psi
\end{eqnarray}
so a left-chiral spinor has nonzero upper components and a right-chiral spinor
has nonzero lower components.  Chirality is conserved in gauge interactions
because the matter Lagrangian, Eq.~(\ref{Lmatter}), connects fields of the same
chirality.  In the massless limit, helicity and chirality are related, because
the factor $\sqrt{E-{\rm p}}$ vanishes in the expressions for the spinors in
Table~\ref{tab:spinors}, causing either the upper or lower components to
vanish:
\begin{eqnarray}
&&u_+(p)=u_R(p)\;\;\;\;u_-(p)=u_L(p)\nonumber\\
&&v_+(p)=v_L(p)\;\;\;\;v_-(p)=v_R(p)
\end{eqnarray}
Note that the relationship between helicity and chirality is reversed for
fermions and antifermions.

\begin{figure}[t]
\begin{center}
\epsfxsize=1.3in \epsfbox{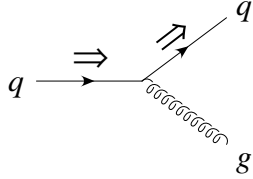}
\end{center}
\caption{Helicity is conserved for massless quarks involved in a gauge
interaction.} \label{helicitycons}
\end{figure}

\begin{figure}[b]
\begin{center}
\begin{picture}(250,90)
\put(0,0){\epsfxsize=3.2in \epsfbox{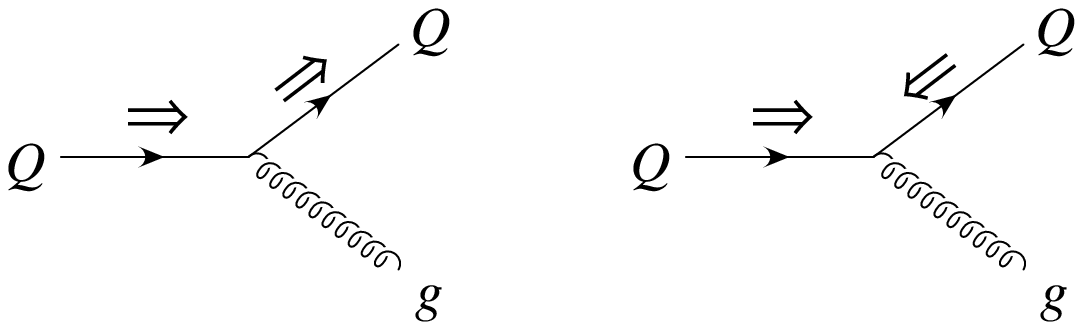}} \put(230,41){\large
$\sim \;\; m$}
\end{picture}
\end{center}
\caption{For massive quarks, there are helicity-conserving and nonconserving
gauge interactions.  The amplitude for the latter is proportional to the quark
mass.} \label{helicityviolation}
\end{figure}

For massless fermions, chirality conservation implies helicity conservation,
as shown in Fig.~\ref{helicitycons}.  For massive fermions, helicity is no
longer related to chirality, so although chirality is conserved, helicity is
not. This is illustrated in Fig.~\ref{helicityviolation}.  Both
helicity-conserving and helicity-nonconserving gauge interactions occur; the
latter are proportional to the fermion mass, since they are forbidden in the
massless limit.
\\[7pt]
{\em Exercise 3.3 ($\ast$) - Do the quark mass terms in ${\cal L}_M$,
Eq.~(\ref{Lmass}), conserve chirality?}
\\[7pt]
\indent A useful discrete symmetry of QCD is parity, ${\bf x}\to -{\bf x}$,
${\bf p}\to -{\bf p}$.  Helicity flips under parity, because although spin
does not flip,\footnote{Spin angular momentum, like orbital angular momentum
($\bf L=\bf x\times \bf p$), does not change sign under parity.} the direction
of motion of the fermion does. One can show that the spinors of
Table~\ref{tab:spinors} are related to each other under parity as follows:
\begin{eqnarray}
&&u_-(p)=\gamma^0u_+(\tilde p)\nonumber \\
&&v_-(p)=\gamma^0v_+(\tilde p)
\end{eqnarray}
where $p=(E,\bf p)$, $\tilde p=(E,-\bf p)$.  This demonstrates that parity
flips the helicity.

\begin{figure}[t]
\begin{center}
\epsfxsize=4.7in \epsfbox{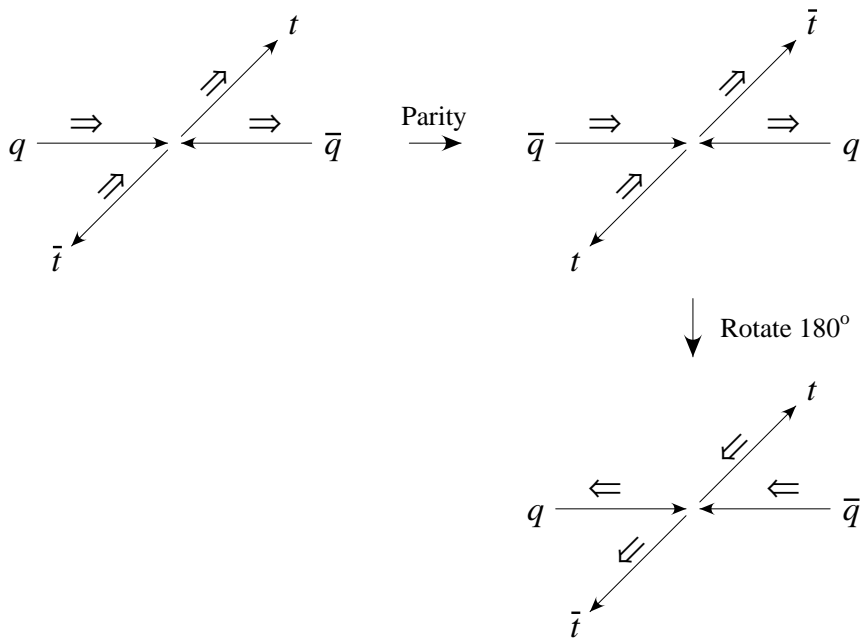}
\end{center}
\caption{Parity and rotational symmetry are used to show that the top quark is
produced unpolarized in (unpolarized) $p\bar p$ collisions.} \label{parity}
\end{figure}

Parity can be used to show that top quarks are produced unpolarized in QCD
reactions.  Let's consider the quark-antiquark-annihilation subprocess, for
example; a similar argument can be given for the gluon-fusion subprocess. In
Fig.~\ref{parity} I show a quark and an antiquark of opposite helicity
annihilating to produce a top quark and a top antiquark of opposite helicity.
(Due to helicity conservation in the massless limit, the helicities of the
light quark and antiquark must be opposite; this is not true of the top quark
and antiquark.)  Applying a parity transformation to this reaction yields the
second diagram in Fig.~\ref{parity}.  Rotating this figure by $180^\circ$ in
the scattering plane yields the third diagram, which is the same as the first
diagram but with all helicities reversed. Since parity is a symmetry of QCD,
the rates for the first and third reactions are the same.  The light quarks are
unpolarized in (unpolarized) $p\bar p$ collisions, so the first and third
reactions will occur with equal probabilities.  The first reaction produces
positive-helicity top quarks, the second negative-helicity top quarks. Thus
top quarks are produced with positive and negative helicity with equal
probability, {\it i.e.}, they are produced unpolarized.

\begin{figure}[t]
\begin{center}
\epsfxsize=4.7in \epsfbox{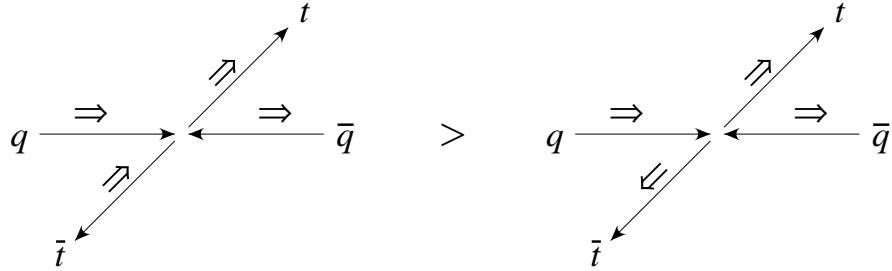}
\end{center}
\caption{The cross section for opposite-helicity $t\bar t$ production is
greater than that for same-helicity $t\bar t$ production.} \label{spincorrel}
\end{figure}

\begin{figure}
\begin{center}
\epsfxsize=2.3in \epsfbox{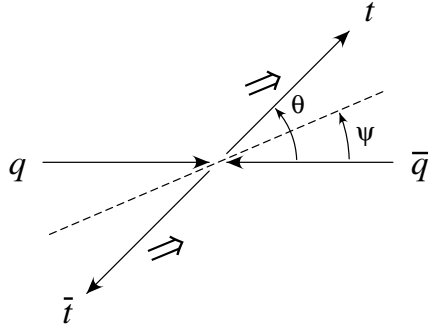}
\end{center}
\caption{The ``off-diagonal'' basis.  The spins of the top quark and antiquark
point in the same direction.} \label{offdiagonal}
\end{figure}

However, there is another avenue open to observe the spin of the top quark.
Although the top quark is produced unpolarized, the spin of the top quark is
correlated with that of the top antiquark.  This is shown in
Fig.~\ref{spincorrel}; the rate for opposite-helicity $t\bar t$ production is
greater than that of same-helicity $t\bar t$ production.
\\[7pt]
{\em Exercise 3.4 ($\ast$) - Argue that in the limit $E\gg m$, the correlation
between the helicities of the top quark and antiquark is $100\%$.}
\\[7pt]
\indent There is a special basis in which the correlation is $100\%$ for all
energies, dubbed the ``off-diagonal'' basis \cite{Mahlon:1997uc}.  This basis
is shown in Fig.~\ref{offdiagonal}. Rather than using the direction of motion
of the quarks as the spin quantization axis, one uses another direction, which
makes an angle $\psi$ with respect to the beam, related to the scattering
angle $\theta$ by
\begin{equation}
\tan\psi=\frac{\beta^2\sin\theta\cos\theta}{1-\beta^2\sin^2\theta}\;,\label{psi}
\end{equation}
where $\beta$ is the velocity of the top quark and antiquark in the
center-of-momentum frame.  When the spin is projected along this axis, the
correlation is $100\%$; the spins of the top quark and antiquark point in the
same direction along this axis.

The moral of this story is that, for massive fermions, there is nothing special
about the helicity basis.  We will see this again in the next section on the
weak interaction.  The spin correlation between top quarks and antiquarks
should be observed for the first time in Run II of the Tevatron.
\\[7pt]
{\em Exercise 3.5 ($\ast$) - Use Eq.~(\ref{psi}) to show that in the limit
$E\gg m$, the off-diagonal basis is identical to the helicity basis.  Argue
that this had to be the case. [Hint: See Exercise 3.4.]}
\\[7pt]
{\em Exercise 3.6 ($\ast\ast\ast$) - What is the off-diagonal basis at
threshold $(E=m)$?  Give a physics argument for this basis at threshold.}

\section{Top Weak Interactions}\label{sec:weak}

\begin{figure}[t]
\begin{center}
\begin{picture}(220,110)
\put(0,0){\epsfxsize=1.3in \epsfbox{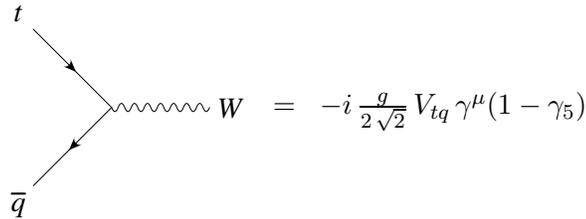}} \put(105,47.5){$= \;\;
-i \, \frac{g}{2\,\sqrt{2}} \, V_{tq} \, \gamma^{\mu}
                         (1 - \gamma_5)$}
\end{picture}
\end{center}
\caption{Charged-current weak interaction of the top quark.}
\label{chargedcurrent}
\end{figure}

In this section I discuss the charged-current weak interaction of the top
quark, shown in Fig.~\ref{chargedcurrent}.  This interaction connects the top
quark with a down-type quark, with an amplitude proportional to the CKM matrix
element $V_{tq}$ ($q=d,s,b$).  The interaction has a vector-minus-axial-vector
($V-A$) structure because only the left-chiral component of the top quark
participates in the $SU(2)$ gauge interaction (see Table~\ref{tab:sm}).

\begin{figure}[b]
\begin{center}
\epsfxsize=1.3in \epsfbox{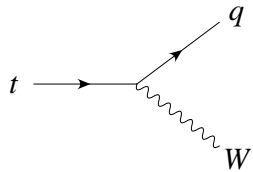}
\end{center}
\caption{Top-quark decay via the charged-current weak interaction.}
\label{topdecay}
\end{figure}

The charged-current weak interaction is responsible for the rapid decay of the
top quark, as shown in Fig.~\ref{topdecay}.  The partial width into the final
state $Wq$ is proportional to $|V_{tq}|^2$.\footnote{The $W$ boson then goes
on to decay to a fermion-antifermion pair.}  The CDF Collaboration has
measured \cite{Affolder:2000xb}
\begin{eqnarray}
&&\frac{BR(t\to Wb)}{BR(t\to Wq)}=0.94^{+0.31}_{-0.24}\nonumber \\
&&=\frac{|V_{tb}|^2}{|V_{td}|^2+|V_{ts}|^2+|V_{tb}|^2}\label{unitarity}
\end{eqnarray}
This implies that $|V_{tb}|\gg |V_{td}|,|V_{ts}|$, but it does not tell us the
absolute magnitude of $V_{tb}$.
\\[7pt]
{\em Exercise 4.1 ($\ast$) - Show that the denominator of the last expression
in Eq.~(\ref{unitarity}) is unity if one assumes that there are just three
generations.}
\\[7pt]
Thus, if we assume three generations, Eq.~(\ref{unitarity}) implies
$|V_{tb}|=0.97^{+0.16}_{-0.12}$.  However, we already know
$V_{tb}=0.9990-0.9993$ if there are just three generations \cite{Hagiwara:pw}.

\begin{figure}[t]
\begin{center}
\epsfxsize=4.7in \epsfbox{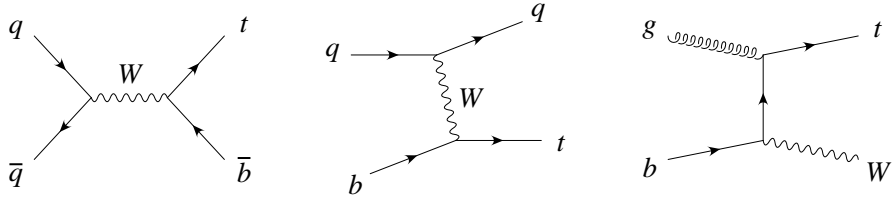}
\end{center}
\caption{Single-top-quark production via the weak interaction.  The first
diagram corresponds to the $s$-channel subprocess, the second to the
$t$-channel subprocess, and the third to $Wt$ associated production (only one
of the two contributing diagrams is shown).} \label{singletop}
\end{figure}

{\em Single top} -- The magnitude of $V_{tb}$ can be extracted directly by
measuring the cross section for top-quark production via the weak
interaction.  There are three such processes, depicted in
Fig.~\ref{singletop}, all of which result in a single top quark rather than a
$t\bar t$ pair \cite{Stelzer:1998ni}.  The cross sections for these single-top
processes are proportional to $|V_{tb}|^2$.

The first subprocess in Fig.~\ref{singletop}, which is mediated by the exchange
of an $s$-channel $W$ boson, is analogous to the Drell-Yan subprocess.  The
second subprocess is simply the first subprocess turned on its side, so the $W$
boson is in the $t$ channel.  The $b$ quark is now in the initial state, so
this subprocess relies on the $b$ distribution function in the proton, which
we will discuss momentarily.\footnote{If one instead uses a $d$ or $s$ quark
in the initial state, the cross section is much less due to the CKM
suppression.} In the third subprocess, the $W$ boson is real, and is produced
in association with the top quark. This subprocess is also initiated by a $b$
quark.  The $s$- and $t$-channel subprocesses should be observed for the first
time in Run II of the Tevatron; associated production of $W$ and $t$ must
await the LHC.

\begin{table}[b]
\caption{Cross sections (pb), at next-to-leading-order in QCD, for top-quark
production via the weak interaction at the Tevatron and the LHC
\cite{Smith:1996ij,Stelzer:1997ns,Zhu:uj}.}
\begin{center}\begin{tabular}[7]{c|c|c|c}
\hline\hline
&$s$ channel&$t$ channel&$Wt$\\
\hline
Tevatron ($\sqrt s=2.0$ TeV $p\bar p$)&$0.90\pm 5\%$&$2.1\pm 5\%$&$0.1\pm 10\%$ \\
LHC ($\sqrt s=14$ TeV $pp$)&$10.6\pm 5\%$&$250\pm 5\%$&$75\pm 10\%$\\
\hline\hline
\end{tabular}\end{center} \label{tab:singletop}
\end{table}

The cross sections for these three single-top processes are given in
Table~\ref{tab:singletop} at the Tevatron and the LHC.  The largest cross
section at both machines is from the $t$-channel subprocess; it is nearly one
third of the cross section for $t\bar t$ pair production via the strong
interaction (see Table~\ref{tab:topsigma}). The next largest cross section at
the Tevatron is from the $s$-channel subprocess.  This is the smallest of the
three at the LHC, because it is initiated by a quark-antiquark collision.  As
is evident from Fig.~\ref{pdfs}, the light-quark distribution functions grow
with decreasing $x$ more slowly than the gluon or $b$ distribution functions,
so quark-antiquark annihilation is relatively suppressed at the LHC.  For a
similar reason, associated production of $W$ and $t$ (which is initiated by a
gluon-$b$ collision) is relatively large at the LHC, while it is very small at
the Tevatron.

\begin{figure}[t]
\begin{center}
\epsfxsize=1.5in \epsfbox{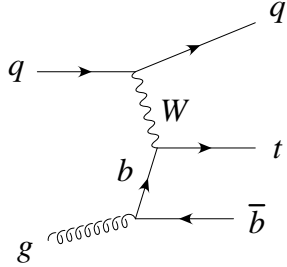}
\end{center}
\caption{When the $\bar b$ is produced at high transverse momentum, the
leading-order process for $t$-channel single-top production is $W$-gluon
fusion.} \label{wgfusion}
\end{figure}

\begin{figure}[t]
\begin{center}
\epsfxsize=3in \epsfbox{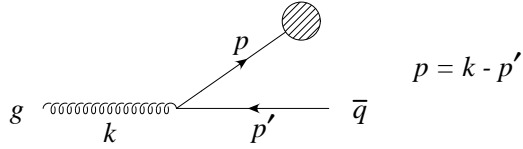}
\end{center}
\caption{When a gluon splits into a real antiquark and a virtual quark, the
quark propagator becomes singular when the kinematics are collinear.}
\label{collineardiv}
\end{figure}

Let's consider the largest of the three processes, $t$-channel single-top
production, in more detail.  This process was originally dubbed $W$-gluon
fusion \cite{Willenbrock:cr}, because it was thought of as a virtual $W$
striking a gluon to produce a $t\bar b$ pair, as shown in
Fig.~\ref{wgfusion}.  If the $\bar b$ in the final state is at high transverse
momentum ($p_T$), this is indeed the leading-order diagram for this process.
If we instead integrate over the $p_T$ of the $\bar b$, we obtain an
enhancement from the region where the $\bar b$ is at low $p_T$, nearly
collinear with the incident gluon.
\\[7pt]
{\em Exercise 4.2 ($\ast\ast$) - Show that a massless quark propagator blows up
in the collinear limit, as shown in Fig.~\ref{collineardiv}.}
\\[7pt]
The $b$ mass regulates the collinear divergence, such that the resulting cross
section is proportional to $\alpha_S\ln (m_t^2/m_b^2)$, where the weak
couplings are tacit.

\begin{figure}[b]
\begin{center}
\epsfxsize=1.5in \epsfbox{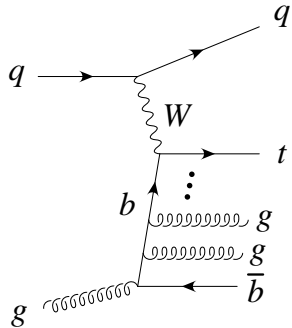}
\end{center}
\caption{The emission of collinear gluons is suppressed only by $\alpha_S\ln
(m_t^2/m_b^2)$, rather than $\alpha_S$.} \label{multiplegluon}
\end{figure}

This collinear enhancement is desirable --- it yields a larger cross section
--- but it also makes perturbation theory less convergent.  Each emission of a
collinear gluon off the internal $b$ quark produces another power of
$\alpha_S\ln (m_t^2/m_b^2)$, because it yields another $b$ propagator that is
nearly on-shell, as shown in Fig.~\ref{multiplegluon}.  The result is that the
expansion parameter for perturbation theory is $\alpha_S\ln (m_t^2/m_b^2)$,
rather than $\alpha_S$ \cite{Stelzer:1997ns}.

Fortunately, there is a simple solution to this problem.  The collinear
logarithms that arise are exactly the ones that are summed to all orders by
the Dokshitzer-Gribov-Lipatov-Altarelli-Parisi (DGLAP) equations.  In order to
sum these logarithms, one introduces a $b$ distribution function in the proton.
When one calculates $t$-channel single-top production using a $b$ distribution
function, as in the second diagram in Fig.~\ref{singletop}, one is
automatically summing these logarithms to all orders.  The expansion parameter
for perturbation theory is now simply $\alpha_S$ \cite{Aivazis:1993pi}.

\begin{figure}[t]
\begin{center}
\epsfxsize=1.8in \epsfbox{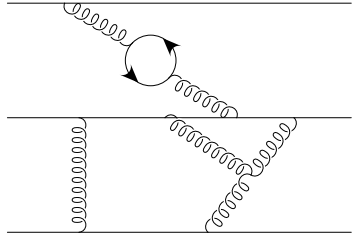}
\end{center}
\caption{The quark ``sea'' in the proton arises from loops of virtual quarks.}
\label{seaquarks}
\end{figure}

Figure~\ref{seaquarks} shows how the $b$ distribution function in the proton
arises from a gluon splitting into a (virtual) $b\bar b$ pair.  The strange
and charm distributions arise in the same way; this also explains the presence
of up and down antiquarks in the proton (see Fig.~\ref{pdfs}).  Unlike the
other ``sea'' quark distributions, which are extracted from experiment, the $b$
distribution function is calculated from the initial condition $b(x)=0$ at
$\mu=m_b$, and is evolved to higher $\mu$ via the DGLAP equations.
\\[7pt]
{\em Exercise 4.3 ($\ast\ast$) - Draw the leading-order Feynman diagrams for
the subprocesses that contribute to
\begin{eqnarray*}
&&{\rm (a)}~p\bar p\to W+X \\
&&{\rm (b)}~p\bar p\to W+1~{\rm jet}+X
\end{eqnarray*}
where $X$ denotes the remnants of the proton and antiproton.}
\\[7pt]
\indent {\em Top-quark spin} -- In the previous section we studied the
top-quark spin in the context of the strong interaction.  Let's now consider
this topic in relation to the weak interaction, beginning with the decay of
the top quark.

\begin{figure}[b]
\begin{center}
\epsfxsize=1.5in \epsfbox{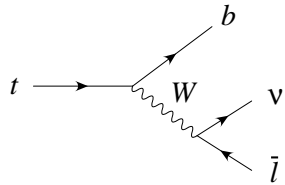}
\end{center}
\caption{Semileptonic top-quark decay.} \label{topsemileptonic}
\end{figure}

The top-quark decay to the final state $b\bar\ell\nu$ is depicted in
Fig.~\ref{topsemileptonic}.
\\[7pt]
{\em Exercise 4.4 ($\ast\ast$) Determine the helicities of all final-state
particles in top decay (neglecting their masses).}
\\[7pt]
The partial width for this decay, summed over the two spin states of the top
quark, is given by a very simple formula:
\begin{equation}
d\Gamma \sim \sum_{\rm spin}|{\cal M}|^2\sim t\cdot\ell b\cdot
\nu\;,\label{width}
\end{equation}
where the four-momentum of the fermion or antifermion is denoted by its label.
To undo the sum over the top-quark spin, it is useful to decompose the
four-momentum of the top quark, $t$, into two lightlike four vectors,
\begin{eqnarray}
&&t=t_1+t_2 \\
&&t_1=\frac{1}{2}(t+ms) \\
&&t_1=\frac{1}{2}(t-ms)
\end{eqnarray}
where $s$ is the spin four-vector.  In the top-quark rest frame, the spin
four-vector is $s=(0,\hat{\bf s})$, where $\hat{\bf s}$ is a unit vector that
defines the spin quantization axis of the top quark.
\\[7pt]
{\em Exercise 4.5 ($\ast$) - Show that $t_1$ and $t_2$ are lightlike
four-vectors, $t_1^2=t_2^2=0$.}
\\[7pt]
\indent In the top-quark rest frame, the spatial components of $t_1$ point in
the spin-up direction, while the spatial components of $t_2$ point in the
spin-down direction. The partial widths for the decay of these two spin states
are
\begin{eqnarray}
&&d\Gamma_{\uparrow}\sim t_2\cdot \ell b\cdot\nu \nonumber\\
&&d\Gamma_{\downarrow}\sim t_1\cdot \ell b\cdot\nu \;.\label{widthspin}
\end{eqnarray}
Note that Eq.~(\ref{width}) is the sum of these two partial widths, as
expected.

\begin{figure}[b]
\begin{center}
\epsfxsize=1.3in \epsfbox{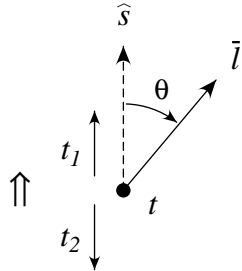}
\end{center}
\caption{Semileptonic top-quark decay in the top rest frame.  The vector $\hat
s$ indicates the spin-quantization axis.  The four-vectors $t_1$ and $t_2$ have
spatial components that point in the spin up and down directions,
respectively.} \label{topangular}
\end{figure}

Let's consider the decay of a top quark with spin up along the $\hat{\bf s}$
direction in its rest frame, as depicted in Fig.~\ref{topangular}.  In this
frame, the spatial components of $t_2$ point in the $-\hat{\bf s}$ direction.
Hence
\begin{equation}
d\Gamma_{\uparrow}\sim t_2\cdot\ell \sim 1+\cos\theta\;,\label{dgamma}
\end{equation}
where $\theta$ is the angle between the spin direction and the charged-lepton
three-momentum (see Fig.~\ref{topangular}).
\\[7pt]
{\em Exercise 4.6 ($\ast$) - Confirm Eq.~(\ref{dgamma}).}
\\[7pt]
Thus
\begin{equation}
\frac{d\Gamma_{\uparrow}}{d\cos\theta}\sim 1+\cos\theta\;,\label{tdecay}
\end{equation}
which means that the charged lepton in top decay tends to go in the direction
of the top-quark spin.  In fact, the charged lepton is the most efficient
analyzer of the top-quark spin, via the angular distribution of
Eq.~(\ref{tdecay}) \cite{Jezabek:1988ja}.

\begin{figure}
\begin{center}
\epsfxsize=1.2in \epsfbox{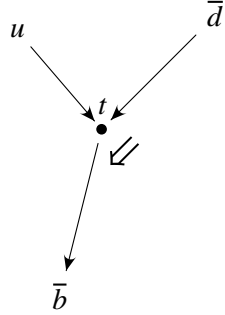}
\end{center}
\caption{In single-top production, the top-quark spin is polarized along the
direction of the $\bar d$ quark in the top rest frame.} \label{toppolarization}
\end{figure}

We can use these same formulas to analyze the top-quark spin in single-top
production \cite{Mahlon:1996pn,Boos:2002xw}.  The Feynman diagram for the
$s$-channel subprocess, Fig.~\ref{singletop}, is the same as that for
top-quark decay, Fig.~\ref{topsemileptonic}, with the replacement $\nu\to u$,
$\bar\ell\to \bar d$. Thus
\begin{eqnarray}
&&d\sigma_{\uparrow}\sim t_2\cdot d b\cdot u \nonumber\\
&&d\sigma_{\downarrow}\sim t_1\cdot d b\cdot u
\end{eqnarray}
from Eqs.~(\ref{widthspin}).  If we choose the spin-quantization axis to point
in the direction of the $\bar d$ (in the top-quark rest frame), then $t_1 \sim
d$, and the latter cross section above vanishes.  Thus the top-quark is $100\%$
polarized in the direction of the $\bar d$ (in the top-quark rest frame) in
$s$-channel single-top production, as depicted in Fig.~\ref{toppolarization}.
This result holds true for $t$-channel single-top production as well, since it
proceeds via the same Feynman diagram, just turned on its side.

Although the top quark is $100\%$ polarized when produced via the weak
interaction, it is not in a state of definite helicity.  Just as we saw in the
previous section, there is nothing special about helicity for massive
fermions.  It may be possible to observe the polarization of single top quarks
in Run II of the Tevatron.
\\[7pt]
{\em Exercise 4.7 ($\ast\ast\ast$) - Show that in the limit $E\gg m$, the top
quark has negative helicity when produced via the $s$- or $t$-channel
subprocesses, as expected.}

\begin{acknowledgments}

I would like to thank Harrison Prosper for organizing a very memorable summer
school.  I am also grateful to Reinhard Schwienhorst and Martin Hennecke for
reading the manuscript, and to Kevin Paul for preparing the figures.  This
work was supported in part by the U.~S.~Department of Energy under contract
No.~DOE~DE-FG02-91ER40677.

\end{acknowledgments}

\section*{Solutions to the exercises}

\subsection*{Section \ref{sec:sm}}

{\em Exercise 1.1} -- It is easiest to show this using index-free notation.
Write the first term in the Lagrangian of Eq.~(\ref{Lmatter}) as
\begin{equation}
i\bar Q_L^i\not\!\!DQ_L^i = iQ_L^\dagger\gamma^0\not\!\!DQ_L\;,
\end{equation}
where $Q_L$ is a 3-component vector in generation space.  This term is
invariant under the transformation $Q_L\to U_{Q_L}Q_L$, because the $3\times
3$ unitary matrix $U_{Q_L}$ commutes with the Dirac matrices (which are the
same for all three generations):
\begin{equation}
iQ_L^\dagger\gamma^0\not\!\!DQ_L\to iQ_L^\dagger
U_{Q_L}^\dagger\gamma^0\not\!\!DU_{Q_L}Q_L=iQ_L^\dagger\gamma^0\not\!\!DQ_L\;,
\end{equation}
where I have used $U_{Q_L}^\dagger U_{Q_L}=1$.  The same argument applies to
the other terms in the matter Lagrangian and their corresponding symmetries.
\\[7pt]
\indent {\em Exercise 1.2} -- Consider the transformation of the first term in
the Yukawa Lagrangian, Eq.~(\ref{LYukawa}), under the symmetry $U_{Q_L}$ of
Eq.~(\ref{U(3)5}):
\begin{equation}
\bar Q_L^i\epsilon \phi^*u_R^j \to \bar Q_L^iU_{Q_L}^\dagger\epsilon
\phi^*u_R^j\;.
\end{equation}
This is not invariant under the symmetry transformation, so $U_{Q_L}$ is
violated.  In contrast, baryon number symmetry, Eq.~(\ref{baryon}), is
respected:
\begin{equation}
\bar Q_L^i\epsilon \phi^*u_R^j \to \bar Q_L^ie^{-i\theta/3}\epsilon
\phi^*e^{i\theta/3}u_R^j=\bar Q_L^i\epsilon \phi^*u_R^j\;.
\end{equation}
The same applies to the other terms in the Yukawa Lagrangian, and also to
lepton number, Eq.~(\ref{lepton}).
\\[7pt]
\indent{\em Exercise 1.3} -- Consider the transformation of the Lagrangian of
Eq.~(\ref{KE}) under the first field redefinition of Eq.~(\ref{fieldredef})
(using the index-free notation introduced in the solution to Exercise 1.1):
\begin{equation}
{\cal L}_{KE} = i\bar u_L\not\!\!\partial u_L\to i\bar
u_LA_{u_L}^\dagger\not\!\!\partial A_{u_L}u_L = i\bar u_L\not\!\!\partial
u_L\;.
\end{equation}
The last step requires that $A_{u_L}$ be unitary, $A_{u_L}^\dagger
A_{u_L}=1$.  The same argument applies to the other fermionic kinetic-energy
terms in the Lagrangian.
\\[7pt]
\indent{\em Exercise 1.4} -- If $A_{u_L}=A_{d_L}$, then we may combine the
first two field redefinitions in Eq.~(\ref{fieldredef}) into one equation:
\begin{equation}
Q_L^i = A_{Q_L}^{ij}Q_L'^j\;,
\end{equation}
where $A_{Q_L}=A_{u_L}=A_{d_L}$. This is exactly the symmetry $U_{Q_L}$ of
Eq.~(\ref{U(3)5}).  The field redefinitions of $u_R^i$ and $d_R^i$ in
Eq.~(\ref{fieldredef}) are the symmetries $U_{u_R}$ and $U_{d_R}$ of
Eq.~(\ref{U(3)5}).
\\[7pt]
\indent{\em Exercise 1.5} -- This follows from the definition of the CKM
matrix, $V\equiv A_{d_L}^\dagger A_{u_L}$:
\begin{equation}
V^\dagger V=(A_{d_L}^\dagger A_{u_L})^\dagger A_{d_L}^\dagger
A_{u_L}=A_{u_L}^\dagger A_{d_L}A_{d_L}^\dagger A_{u_L}=1\;,
\end{equation}
where I have used the unitarity of the $A$ matrices.
\\[7pt]
\indent{\em Exercise 1.6} -- A useful equation to remember is $\hbar c = 197$
MeV fm. Using this, one can convert length to ${\rm mass}^{-1}$:
\begin{equation}
{\rm length} = \hbar c/{\rm mass}\,c^2 = {\rm mass}^{-1}
\end{equation}
using $\hbar = c = 1$.
\\[7pt]
\indent{\em Exercise 1.7} -- Such a term is not invariant under SU(3) gauge
symmetry, $Q_L \to UQ_L$, where $U$ acts on the (suppressed) color indices of
the quarks:
\begin{equation}
Q_L^{iT}\epsilon\phi C \phi^T\epsilon Q_L^j \to Q_L^{iT}U^T\epsilon\phi C
\phi^T\epsilon UQ_L^j\;.
\end{equation}
This involves $U^TU$, which is not equal to unity (rather, $U^\dagger U=1$).
This term is also not invariant under $U(1)_Y$, as the total hypercharge is
nonzero (1/6+1/6+1/2+1/2).
\\[7pt]
\indent{\em Exercise 1.8} -- Lepton number, Eq.~(\ref{lepton}), is violated
because
\begin{equation}
L_L^{iT}\epsilon\phi C \phi^T\epsilon L_L^j\to L_L^{iT}e^{i\phi}\epsilon\phi C
\phi^T\epsilon e^{i\phi}L_L^j
\end{equation}
is not invariant.  Recall that lepton number is an accidental symmetry of the
standard model.  Once you go beyond the standard model by including
higher-dimension operators, there is no reason for lepton number (and baryon
number) to be conserved.
\\[7pt]
\indent{\em Exercise 1.9} -- We'll follow a similar argument as the one made to
count the number of parameters in the CKM matrix.  The Yukawa matrix
$\Gamma_e$ has $2\times 3\times 3$ parameters, and the complex, symmetric
matrix $c^{ij}$ has $2\times 6$ parameters.  The symmetries $U_{L_L}$ and
$U_{e_R}$ contain $2\times 3\times 3$ degrees of freedom, so the number of
physically-relevant parameters is
\begin{equation}
2\times 3\times 3 + 2\times 6 - 2\times 3\times 3 = 12\;.
\end{equation}
[Note that we did not remove lepton number from the symmetries, because lepton
number is violated by ${\cal L}_5$, Eq.~(\ref{L5})].  Of these parameters, six
are the charged-lepton and neutrino masses, leaving six parameters for the MNS
matrix.  Three are mixing angles, and three are $CP$-violating phases.

\subsection*{Section \ref{sec:virtual}}

\indent{\em Exercise 2.1} -- Plug the expressions for $\alpha$, $G_F$, and
$M_Z$ in terms of $g$, $g'$, and $v$, given at beginning of Section
\ref{sec:virtual}, into Eq.~(\ref{MW}) and carry through the algebra to obtain
$M_W^2 = (1/4)g^2v^2$.
\\[7pt]
\indent{\em Exercise 2.2} -- Using Eq.~(\ref{s2w}), we can write
Eq.~(\ref{MW2}) as
\begin{equation}
M_W^2\left(1-\frac{M_W^2}{M_Z^2}\right)=\frac{\pi\alpha}{\sqrt
2G_F}\;.\label{MW3}
\end{equation}
Solving this quadratic equation for $M_W^2$ yields Eq.~(\ref{MW}).
Alternatively, one could plug the expressions for $\alpha$, $G_F$, and $M_Z$
in terms of $g$, $g'$, and $v$, given at beginning of Section
\ref{sec:virtual}, as well as $M_W^2=(1/4)g^2v^2$, into the above equation to
check its veracity.
\\[7pt]
\indent{\em Exercise 2.3} -- Starting from Eq.~(\ref{MWloop}), the one-loop
analogue of Eq.~(\ref{MW3}) is
\begin{equation}
M_W^2\left(1-\frac{M_W^2}{M_Z^2}\right)=\frac{\frac{\pi\alpha}{\sqrt
2G_F}}{(1-\Delta r)}\;.
\end{equation}
The differential of this equation (with respect to $M_W^2$ and $m_t^2$,
keeping everything else fixed) is
\begin{equation}
dM_W^2-2\frac{M_W^2}{M_Z^2}dM_W^2=-\frac{\frac{\pi\alpha}{\sqrt
2G_F}}{(1-\Delta r)^2}\frac{3G_Fdm_t^2}{8\sqrt
2\pi^2}\frac{1}{t_W^2}\;,\label{diff}
\end{equation}
where I have used Eq.~(\ref{deltartop}) for $\Delta r$.  We can now set
$\Delta r=0$ to leading-order accuracy, and solve for $dM_W^2/dm_t^2$:
\begin{equation}
\frac{dM_W^2}{dm_t^2}=\frac{3\alpha}{16\pi}\frac{1}{(2c^2_W-1)t^2_W}\;,\label{dMWdmt}
\end{equation}
where I've used Eq.~(\ref{s2w}).  Using $dM_W/dm_t = (m_t/M_W)dM_W^2/dm_t^2$
and evaluating numerically (for $M_W=80$ GeV, $m_t=175$ GeV) gives a slope of
0.0060, in good agreement with the slope of the lines of constant Higgs mass
in Fig.~\ref{mwmt}.
\\[7pt]
\indent{\em Exercise 2.4} -- The desired result follow from inserting
$M_W^2=(1/4)g^2v^2$ and $M_Z^2=(1/4)(g^2+g'^2)v^2$ [Eq.~(\ref{Wmass})] into
the on-shell definition of $\sin^2\theta_W$, Eq.~(\ref{s2w}).
\\[7pt]
\indent{\em Exercise 2.5} -- Combining Eqs.~(\ref{MSbar}) and (\ref{rho}) to
eliminate $\hat s^2_Z$ gives
\begin{equation}
M_W^2\left(1-\frac{M_W^2}{M_Z^2\hat\rho}\right)=\frac{\frac{\pi\alpha}{\sqrt
2G_F}}{(1-\Delta\hat r_W)}\;.
\end{equation}
The differential of this equation is
\begin{equation}
dM_W^2-2\frac{M_W^2}{M_Z^2\hat\rho}dM_W^2+\frac{M_W^4}{M_Z^2\hat\rho^2}
\frac{3G_Fdm_t^2}{8\sqrt 2\pi^2}=0\;,\label{diffMSbar}
\end{equation}
where I have used Eq.~(\ref{rho}) for $\hat\rho$ (there is no $m_t$ dependence
in $\Delta\hat r_W$).  We can now set $\hat\rho=1$ to leading-order accuracy.
Using the leading-order expressions of Eq.~(\ref{MW2}) and $M_W^2/M_Z^2=c^2_W$,
it is easy to show that Eq.~(\ref{diffMSbar}) is identical Eq.~(\ref{diff}) at
leading order.

\subsection*{Section \ref{sec:strong}}

\indent{\em Exercise 3.1} -- The four-momenta of the quark and antiquark in the
center-of-momentum frame are
\begin{eqnarray*}
&&P_1=(E,0,0,p) \\
&&P_2=(E,0,0,-p)\;.
\end{eqnarray*}
Thus $S\equiv (P_1+P_2)^2 = (2E,0,0,0)^2=(2E)^2$, which is the square of the
total energy of the collision.  The last expression in Eq.~(\ref{S}) follows
from $(P_1+P_2)^2=P_1^2+P_2^2+2P_1\cdot P_2\approx 2P_1\cdot P_2$, if we
neglect the proton mass, $P_1^2=P_2^2=m_p^2$.
\\[7pt]
\indent{\em Exercise 3.2} -- Inserting Eq.~(\ref{higgsfield}) into the second
term in the Yukawa Lagrangian, Eq.~(\ref{LYukawa}), yields
\begin{equation}
{\cal L}_{Y} = -\Gamma_d^{ij}\frac{1}{\sqrt 2}(v+h)\bar d_L^i d_R^j + h.c.
\end{equation}
(analogous results are obtained for the other terms in the Lagrangian).  Using
Eq.~(\ref{massmatrix}), this can be written
\begin{equation}
{\cal L}_{Y} = -M_d^{ij}\left(1+\frac{h}{v}\right)\bar d_L^i d_R^j + h.c.
\end{equation}
The field redefinitions that diagonalize the mass matrix,
Eq.~(\ref{fieldredef}), will therefore also diagonalize the couplings of the
fermions to the Higgs boson.  The coupling to a given fermion is thus given by
$-m_f/v$ (times a factor $i$ since the Feynman rules come from $i{\cal L}$),
as shown in Fig.~\ref{higgscoupling}.
\\[7pt]
\indent{\em Exercise 3.3} -- The answer is evidently no, since these terms
connect fields of different chirality.
\\[7pt]
\indent{\em Exercise 3.4} -- In the ultrarelativistic limit, $E\gg m$, the
mass of the top quark is negligible.  Since helicity is conserved for massless
quarks, the top quark and antiquark must be produced with opposite helicities.
\\[7pt]
\indent{\em Exercise 3.5} -- In the limit $E\gg m$ ($\beta\to 1$),
Eq.~(\ref{psi}) implies $\psi=\theta$, which means that the off-diagonal and
helicity bases are the same.  This is as expected, because in the massless
limit the helicities of the top quark and antiquark are $100\%$ correlated
(see Exercise 3.4), which is the defining characteristic of the off-diagonal
basis.
\\[7pt]
\indent{\em Exercise 3.6} -- At threshold ($\beta\to 0$), Eq.~(\ref{psi})
implies $\psi=0$, which means that the top quark and antiquark spins are
$100\%$ correlated along the beam direction.  This is a consequence of
angular-momentum conservation.  At threshold, the top quark and antiquark are
produced at rest with no orbital angular momentum.  The colliding light quark
and antiquark have no orbital angular momentum along the beam direction.  Thus
spin angular momentum along the beam direction must be conserved.  The light
quark and antiquark have opposite helicity (due to helicity conservation in
the massless limit), so the top quark and antiquark are produced with their
spins pointing in the same direction along the beam.

\subsection*{Section \ref{sec:weak}}

\begin{figure}
\begin{center}
\epsfxsize=3in \epsfbox{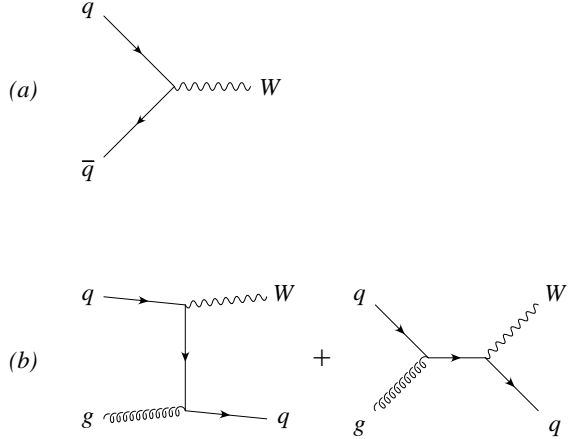}
\end{center}
\caption{(a) Leading-order subprocess for $W$ production. (b) Leading-order
subprocess for $W+1$ jet production.} \label{drellyan}
\end{figure}

\begin{figure}
\begin{center}
\epsfxsize=1.4in \epsfbox{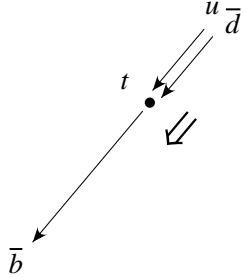}
\end{center}
\caption{Single-top production in the ultrarelativistic limit, as viewed from
the top rest frame.} \label{toppolboost}
\end{figure}

\indent{\em Exercise 4.1} -- This follows from the unitarity of the CKM matrix,
$VV^\dagger=1$. Displaying indices, this may be written
\begin{equation}
V_{ik}V^\dagger_{kj}=V_{ik}V^*_{jk}=\delta_{ij}\;.
\end{equation}
For $i=j$, this implies
\begin{equation}
\sum_{k=d,s,b} |V_{ik}|^2 =1\;,
\end{equation}
which yields the desired result for $i=t$.
\\[7pt]
\indent{\em Exercise 4.2} -- The square of the four-momentum of the quark
propagator in Fig.~\ref{collineardiv} is
\begin{equation}
p^2=(k-p^\prime)^2 = -2k\cdot p^\prime
\end{equation}
This vanishes for collinear kinematics,
\begin{eqnarray*}
&&k=(E,0,0,E)\\
&&p'=(E',0,0,E')\;.
\end{eqnarray*}
Thus the denominator of the quark propagator vanishes in the collinear limit
(if we neglect the quark mass).
\\[7pt]
\indent{\em Exercise 4.3} -- (a) There is just one diagram, shown in
Fig.~\ref{drellyan}(a). (b) There are two contributing subprocesses, $gq\to
Wq$ and $q\bar q\to Wg$; each consists of two Feynman diagrams, shown in
Fig.~\ref{drellyan}(b) for $gq\to Wq$.  The two diagrams for $q\bar q\to Wg$
may be obtained by radiating a gluon off either fermion line in
Fig.~\ref{drellyan}(a).
\\[7pt]
\indent{\em Exercise 4.4} -- The charged-current weak interaction couples only
to left-chiral fields.  Thus the fermions in the final state ($b,\nu$) have
negative helicity, and the antifermion ($\bar\ell$) has positive helicity, due
to the relationship between chirality and helicity for massless particles
(discussed in Section~\ref{sec:strong}).
\\[7pt]
\indent{\em Exercise 4.5} -- In the top-quark rest frame, $s^2=(0,\hat{\bf
s})^2 = -1$, since $\hat{\bf s}$ is a unit vector.  Because $s^2$ is Lorentz
invariant, this is true in all reference frames.  Similarly, $t\cdot s=0$,
because $t=(m,0,0,0)$ in the top-quark rest frame. Thus
\begin{equation}
t_1^2 = \frac{1}{4}(t+ms)^2 = \frac{1}{4}(m^2-m^2+2mt\cdot s) = 0\;,
\end{equation}
and similarly for $t_2^2$.
\\[7pt]
\indent{\em Exercise 4.6} -- The spatial part of the lightlike four-vector
$t_2$ is pointing in the $-\hat{\bf s}$ direction. Thus $t_2\cdot \ell\sim
1-\cos\alpha$, where $\alpha$ is the angle between $-\hat{\bf s}$ and the
direction of the charged lepton.  This angle is supplementary to $\theta$
($\alpha+\theta=\pi$), so $t_2\cdot \ell\sim 1-\cos\alpha=1+\cos\theta$.
\\[7pt]
\indent{\em Exercise 4.7} -- The $s$-channel subprocess, in the top-quark rest
frame, looks like Fig.~\ref{toppolarization}.  In the limit $E\gg m$, this
figure looks like Fig.~\ref{toppolboost}; the $u$ and $\bar d$ approach each
other along a line and annihilate to make a top quark at rest and a $\bar b$
that carries off the incoming momentum.  As always, the top-quark spin points
in the direction of the $\bar d$.  To view this event from the
center-of-momentum frame, one boosts opposite the direction of motion of the
$u$ and $\bar d$. This boosts the top quark in the direction opposite its
spin, so it is in a state of negative helicity. This is as expected; in the
limit $E\gg m$, the top quark acts like a massless quark, and is therefore
produced in a negative-helicity state by the weak interaction (see Exercise
4.4).

\end{document}